\documentclass[fleqn,usenatbib]{mnras}

\usepackage{newtxtext,newtxmath}

\usepackage[T1]{fontenc}

\DeclareRobustCommand{\VAN}[3]{#2}
\let\VANthebibliography\thebibliography
\def\thebibliography{\DeclareRobustCommand{\VAN}[3]{##3}\VANthebibliography}

\usepackage{graphicx}	
\usepackage{amsmath}	

\defcitealias{el-badry_birth_2021}{E21}

\title[Sodium enhancement in evolved CVs]{Sodium enhancement in evolved cataclysmic variables}

\author[]{Natsuko Yamaguchi,$^{1}$\thanks{E-mail: nyamaguc@caltech.edu}
Kareem El-Badry$^{1,2}$, Antonio C. Rodriguez$^{1}$, Maude Gull$^{3}$, Benjamin R. Roulston$^{1}$, \newauthor Zachary P. Vanderbosch$^{1}$
\\
$^{1}$ California Institute of Technology, Department of Astronomy, 
1200 E. California Blvd, Pasadena, CA, 91125, USA \\
$^{2}$Center for Astrophysics $|$ Harvard \& Smithsonian, 60 Garden Street, Cambridge, MA 02138, USA\\
$^{3}$ Department of Astronomy, University of California Berkeley, Berkeley, CA 94720, USA
}

\date{Accepted XXX. Received YYY; in original form ZZZ}
 
\pubyear{2023}

\begin{document}
\label{firstpage}
\pagerange{\pageref{firstpage}--\pageref{lastpage}}
\maketitle

\renewcommand{\arraystretch}{1.1}

\begin{abstract}
We present follow-up spectroscopy of 21 cataclysmic variables (CVs) with evolved secondaries and ongoing or recently-terminated mass transfer. Evolutionary models predict that the secondaries should have anomalous surface abundances owing to nuclear burning in their cores during their main-sequence evolution and subsequent envelope stripping by their companion white dwarfs. To test these models, we measure sodium (Na) abundances of the donors from the Fraunhofer "D" doublet. Accounting for interstellar absorption, we find that {\it all} objects in our sample have enhanced Na abundances. We measure 0.3 $\lesssim$ [Na/H] $\lesssim$ 1.5 dex across the sample, with a median [Na/H] = 0.956 dex, i.e., about an order of magnitude enhancement over solar values. To interpret these values, we run MESA binary evolution models of CVs in which mass transfer begins just as the donor leaves the main sequence. These generically predict Na enhancement in donors with initial donor masses $\gtrsim 1\,M_{\odot}$, consistent with our observations. In the models, Na enrichment occurs in the donors' cores via the NeNa cycle near the end of their main-sequence evolution. Na-enhanced material is exposed when the binaries reach orbital periods of a few hours. Donors with higher initial masses are predicted to have higher Na abundances at fixed orbital period owing to their higher core temperatures during main-sequence evolution. The observed [Na/H] values are on average $\approx$0.3 dex higher than predicted by the models. Surface abundances of evolved CV donors provide a unique opportunity to study nuclear burning products in the cores of intermediate-mass stars. 

\end{abstract}




\section{Introduction}

Cataclysmic variables (CVs) are binary systems in which a non-degenerate star (the ``donor'' or ``secondary'') transfers mass to a white dwarf (WD) companion through stable Roche lobe overflow \citep[see][for a review]{warner_1995}. CV evolution is fundamentally governed by angular momentum loss (AML), which shrinks CV orbits and drives mass transfer. The strength and dominant mechanism of AML in CVs is still uncertain, and varies with orbital period. In the standard evolutionary model \citep[e.g.][]{Knigge2011ApJS}, AML at orbital periods $P_{\rm orb} \gtrsim 3$ h is driven primarily by magnetic braking, while gravitational wave radiation dominates at shorter periods.  Weakening of magnetic braking at $P_{\rm orb}\lesssim 3$ h is proposed to give rise to the CV ``period gap''; i.e. the observed lack of CVs with orbital periods between 2 and 3 hours \citep[e.g.][]{rappaport_new_1983, Knigge2006MNRAS, Knigge2011ApJS, Inight2021MNRAS, Pala2022MNRAS}. 

The vast majority of CV donors fall on a tight ``donor sequence'', in which physical parameters such as mass, radius, temperature, and spectral type depend primarily on orbital period \citep[e.g.][]{Beuermann1998A&A, smith_secondary_1998, Knigge2006MNRAS, Abrahams2020arXiv}. In the standard evolutionary model, CV donors evolve along this sequence  from long to short periods as their donors are whittled down by mass transfer. Objects on the donor sequence have temperatures and radii similar to main-sequence stars of the same mass. Although mass transfer causes some radius inflation relative to main-sequence stars, this inflation is modest because objects on the donor sequence are only mildly out of thermal equilibrium \citep[e.g.][]{Knigge2006MNRAS,  Knigge2011ApJS}.

A number of CVs have been discovered over the years with physical parameters that differ significantly from the standard donor sequence. Observationally, these systems stood out from the bulk of the CV population for having unusually warm and luminous donors compared to normal CVs at the same orbital period \citep[e.g.][]{thorstensen_qz_2002, thorstensen_1rxs_2002, thorstensen_css_2013, rebassa-mansergas_sdss_2014}. Several of the same systems also have infrequent outbursts and faint disks, suggesting low mass transfer rates. Evolutionary models predict that such ``evolved CVs'' can form if mass transfer began when the secondary star was near the end of its main sequence lifetime and thus had a core enhanced in helium \citep[e.g.][]{podsiadlowski_cataclysmic_2003}. At the short periods where they are observed, the donors in these systems have semi-degenerate helium cores and thick hydrogen-burning envelopes. This leads them to have quite different physical parameters from ordinary CV donors, which consist mostly of hydrogen. 


Evolved CVs are predicted to follow qualitatively different evolutionary pathways from ordinary CVs. Their donor’s cores remain radiative (as opposed to donors in normal CVs which become fully convective) so they are not expected to detach at $P_{\rm orb}\approx 3$ hours as a result of weakened magnetic braking. Instead, this can occur at a wide range of orbital periods when the donors lose enough of their convective envelopes at $T_{\rm eff}\gtrsim 6500\,\rm K$. These donors are predicted to detach from their Roche lobes and be observable as extremely low mass (ELM) WDs in close binaries \citep{sun_formation_2018, Li2019, el-badry_birth_2021}.

Evolved CVs provide one possible formation channel for the AM Canum Venaticorum binaries \citep[AM CVns;][]{Tutukov1985SvAL, podsiadlowski_cataclysmic_2003, kalomeni_evolution_2016}, which are ultra-compact binaries with helium donors and $P_{\rm orb}\lesssim 70$ minutes, below the minimum period of normal CVs. AM CVns are an especially interesting outcome of evolved CV evolution since they will be among the loudest sources of gravitational waves for LISA \citep{Breivik2018ApJL, Kupfer2018MNRAS}. 
At longer periods, CVs with evolved donors can have higher mass transfer rates than ordinary CVs and can be observed as supersoft X-ray sources \citep[e.g.][]{Li1997A&A, Schenker2002}. 
These high accretion rates can lead to stable burning on the surface of the accretion WD and may provide one channel for Type Ia (or .Ia) supernovae \citep{livne_successive_1990, bildsten_faint_2007, Nomoto2007ApJ, Wolf2013ApJ, brooks_am_2015}.

Because evolved CV donors completed most of their main-sequence evolution before the onset of mass transfer, their photospheres -- which contain material previously inside their donors' convective cores -- are expected to have unusual surface abundances. The first observed evolved CVs, EI Psc and QZ Ser, bore out this prediction, with optical spectra that hinted at helium (He) and sodium (Na) enhancement \citep{thorstensen_1rxs_2002, thorstensen_qz_2002}. Ultraviolet spectroscopy soon showed that the same systems were enhanced in nitrogen and deficient in carbon, as expected if material on the donors' surface was previously processed by CNO burning \citep{haswell_ultraviolet_2002, Gaensicke2003, Toloza2022}. Studies based on infrared spectroscopy have similarly reported carbon deficits \citep{harrison_abundance_2016, harrison_identification_2018}.

A large sample of CVs with evolved donors was presented by \citet[][hereafter \citetalias{el-badry_birth_2021}]{el-badry_birth_2021}. Using light curves from the Zwicky Transient Facility (ZTF; \citealt{Bellm2019PASP}), they identified 21 binaries with $P_{\rm orb} < 6$ h exhibiting large-amplitude ellipsoidal variability and falling below the main-sequence in the color-magnitude diagram. The donors in these systems have comparable temperatures and surface gravities to main-sequence A and F stars, but significantly lower masses and smaller radii. Using low-resolution follow-up spectroscopy, they showed that some objects in the sample have ongoing mass transfer, while others likely just recently became detached. They noted hints of excess Na absorption for all donors with $T_{\rm eff} \lesssim$ 7000 K, suggesting potential Na enhancement (at higher temperatures, more of the Na is ionised so neutral lines in the optical become weaker).

A literature search indeed reveals several other scattered reports of Na enhancement in the donors of evolved CVs \citep[e.g.][]{thorstensen_1rxs_2002, thorstensen_css_2013, harrison_identification_2018, Green2020, Zhang2022}. There has, however, been little systematic study or quantitative evolutionary predictions of the expected enhancement. In contrast to CNO lines, Na has strong optical lines, namely the 5900\AA \ doublet (i.e. Fraunhofer D-lines, at 5890 and 5896\AA) and the 8200\AA \ doublet (at 8183 and 8195\AA), making it readily accessible in evolved CVs (where the donor, rather than the disk, dominates in the optical). Na is produced in advanced H burning processes which can only take place in the high-temperature interiors of stars with masses $M\gtrsim 1 \ M_{\odot}$, making it a potential diagnostic of the initial masses of the donors in evolved CVs.

In this paper, we follow up on the work of \citetalias{el-badry_birth_2021} and obtain higher-resolution and higher-SNR spectra of their targets to look for and quantify potential excesses in Na absorption lines compared to model spectra. We also use the Modules for Experiments in Stellar Astrophysics (MESA) to model binary evolutionary scenarios that give rise to evolved CVs and test whether they are able to predict the observed abundances. In Section \ref{sec:esi}, we describe our spectroscopic observations as well as the subsequent data reduction and calculations of several parameters of our objects using the spectra. In Section \ref{sec:analysis}, we explain the process of generating a grid of model stellar spectra with different chemical abundances. By comparing equivalent widths of Na lines in the model spectra to our data, we infer the Na abundances of our objects. We then describe and present the results of our MESA models in Section \ref{sec:mesa}. 

\section{Spectroscopic observations} \label{sec:esi}

We observed all 21 evolved CVs from \citetalias{el-badry_birth_2021} across 4 nights. A detailed observing log can be found in Appendix~\ref{appendix:log}.

We observed 20 targets with the Echellette Spectrograph and Imager (ESI; \citealt{Sheinis2002PASP}) on the Keck II telescope in echellette mode. For the majority of these observations, we used the 0.5'' slit with a 600s exposure which yielded spectral resolution $R \sim 9300$ and a typical SNR $\sim$ 20 per pixel. A few targets were observed with other ESI slits; see Table \ref{tab:log} for details. All ESI spectra were reduced using the MAuna Kea Echelle Extraction (MAKEE) pipeline, which performs bias-subtraction, flat fielding, wavelength calibration, sky subtraction, and a flexure correction to the wavelength solution using sky lines. 

Due to poor weather conditions affecting several observing runs and observability constraints during the year, we observed one object, P\_2.74a, with the Low Resolution Imaging Spectrometer (LRIS; \citealt{Oke1995PASP}) on the Keck I telescope. We observed the binary over a full orbit, with 63 (red) and 78 (blue) $\times$ 90 second exposures and a 1.0'' slit (see Appendix~\ref{appendix:log}), but most of the abundance analysis focuses on a single exposure, taken at quadrature. The red and blue arms covered a wavelength range of $\sim$ 3150 to 5600\AA \ and 5400 to 10290\AA, respectively. The data was reduced using LPipe \citep{Perley2019PASP}. This provided a significantly lower resolution of  $R \sim 1540$ compared to the ESI spectra but as we are primarily concerned with the calculation of equivalent widths, this should be acceptable to complete the sample. 

Cutouts of the spectra for all objects can be found in Figures~\ref{fig:mg_triplet},~\ref{fig:EW_regions}, and~\ref{fig:h_alpha}, which respectively show spectral regions containing the Mg I triplet, the Na D doublet, and the H$\alpha$ line.

The phases at which the spectra were taken could be deduced from the recorded mid-exposure time and the orbital ephemeris obtained from light curve fitting, summarized in Table 4 of \citetalias{el-badry_birth_2021}. 

\citetalias{el-badry_birth_2021} also inferred the effective temperatures of all the targets from SED fitting of UV-to-IR photometry from Pan-STARRS, WISE, and 2MASS, using the results from fitting their low-resolution spectra as a weak prior. These temperatures, as well as several other relevant parameters from the same paper, are summarized in Table \ref{tab:BoE_info}.

 \begin{table*} 
     \centering
     \begin{tabular}{|c c c c c c|}
          \hline
          ID & Gaia eDR3 ID & $P_{\rm orb}$ [days] & $T_{\rm eff}$ [K] &  $R_{\rm donor}$ [$R_{\odot}$] & $i$ [deg] \\
          \hline
         P\_2.00a & 4393660804037754752 & 0.08331301(2) & 7734 $\pm$ 73.0 & 0.17$^{+0.01}_{-0.01}$ & 55.49$^{+14.14}_{-5.33}$ \\
         P\_2.74a & 861540207303947776 & 0.11414278(4) & 4726 $\pm$ 52.0 & 0.23$^{+0.01}_{-0.01}$ & 79.76$^{+7.0}_{-6.75}$ \\
         P\_3.03a & 1030236970683510784 & 0.12642666(6) & 5910 $\pm$ 77.0 & 0.27$^{+0.03}_{-0.03}$ & 73.10$^{+11.29}_{-9.92}$ \\
         P\_3.06a & 2133938077063469696 & 0.1274832(2) & 5862 $\pm$ 70.0 & 0.27$^{+0.03}_{-0.02}$ & 79.49$^{+7.09}_{-6.61}$ \\
         P\_3.13a & 4228735155086295552 & 0.13056512(8) & 6662 $\pm$ 54.0 & 0.28$^{+0.01}_{-0.01}$ & 83.66$^{+4.65}_{-2.75}$ \\
         P\_3.21a & 45968897530496384 & 0.13375228(7) & 7022 $\pm$ 72.0 & 0.31$^{+0.02}_{-0.02}$ & 71.56$^{+7.29}_{-3.18}$ \\
         P\_3.43a & 184406722957533440 & 0.1429875(2) & 6193 $\pm$ 83.0 & 0.26$^{+0.01}_{-0.01}$ & 79.21$^{+7.31}_{-6.44}$ \\
         P\_3.48a & 4002359459116052864 & 0.14479504(6) & 6444 $\pm$ 62.0 & 0.26$^{+0.02}_{-0.02}$ & 73.90$^{+11.05}_{-10.57}$ \\
         P\_3.53a & 1965375973804679296 & 0.146883(3) & 5324 $\pm$ 66.0 & 0.32$^{+0.01}_{-0.01}$ & 79.66$^{+6.96}_{-6.67}$ \\
         P\_3.81a & 373857386785825408 & 0.1586339(7) & 6689 $\pm$ 72.0 & 0.35$^{+0.07}_{-0.05}$ & 78.34$^{+7.76}_{-7.71}$ \\
         P\_3.88a & 1077511538271752192 & 0.1615714(3) & 5875 $\pm$ 102.0 & 0.28$^{+0.02}_{-0.01}$ & 79.64$^{+6.96}_{-6.6}$ \\
         P\_3.90a & 3053571840222008192 & 0.1624549(1) & 7442 $\pm$ 87.0 & 0.33$^{+0.01}_{-0.01}$ & 64.85$^{+10.19}_{-4.42}$ \\
         P\_3.98a & 896438328413086336 & 0.16582294(9) & 5122 $\pm$ 53.0 & 0.40$^{+0.03}_{-0.03}$ & 70.36$^{+13.16}_{-12.32}$ \\
         P\_4.06a & 4358250649810243584 & 0.16898413(4) & 7587 $\pm$ 72.0 & 0.28$^{+0.01}_{-0.01}$ & 59.48$^{+12.19}_{-5.3}$ \\
         P\_4.10a & 3064766376117338880 & 0.1708102(4) & 4846 $\pm$ 38.0 & 0.37$^{+0.01}_{-0.01}$ & 79.51$^{+7.05}_{-6.58}$ \\
         P\_4.36a & 2126361067562200320 & 0.181743(2) & 5998 $\pm$ 80.0 & 0.39$^{+0.02}_{-0.02}$ & 74.15$^{+10.74}_{-10.16}$ \\
         P\_4.41a & 2171644870571247872 & 0.1838691(2) & 7013 $\pm$ 105.0 & 0.35$^{+0.01}_{-0.01}$ & 64.02$^{+10.6}_{-3.9}$ \\
         P\_4.47a & 1315840437462118400 & 0.18604299(9) & 7171 $\pm$ 108.0 & 0.36$^{+0.07}_{-0.05}$ & 73.03$^{+6.83}_{-2.77}$ \\
         P\_4.73a & 2006232382792027904 & 0.1969063(2) & 7619 $\pm$ 85.0 & 0.38$^{+0.01}_{-0.01}$ & 57.45$^{+12.3}_{-5.32}$ \\
         P\_5.17a & 4382957882974327040 & 0.2153735(2) & 6187 $\pm$ 61.0 & 0.37$^{+0.03}_{-0.03}$ & 67.57$^{+14.9}_{-11.87}$ \\
         P\_5.42a & 286337708620373632 & 0.2258108(1) & 7326 $\pm$ 78.0 & 0.40$^{+0.01}_{-0.01}$ & 68.34$^{+8.65}_{-4.01}$ \\
          \hline  
     \end{tabular}
     \caption{Physical parameters of the 21 objects, combining information from Tables 1, 2, and 5 of \citetalias{el-badry_birth_2021}. The ``ID''s are from \citetalias{el-badry_birth_2021}, with the numbers corresponding to the orbital period in hours. $P_{\rm orb}$ is the orbital period in days from ZTF light curves and confirmed by spectral fitting; $T_{\rm eff}$ is the effective temperature of the donor from SED fitting; $M_{\rm donor}$, $R_{\rm donor}$, and $i$ is the mass of the donor, its radius, and the minimum inclination respectively, obtained from a combined fitting using the Markov chain Monte Carlo (MCMC) method for properties of both the donor and WD companion.}
     \label{tab:BoE_info}
 \end{table*}

\subsection{Spectra and radial velocities} \label{ssec:rv}

\begin{figure*}
    \centering
    \includegraphics[width=0.95\textwidth]{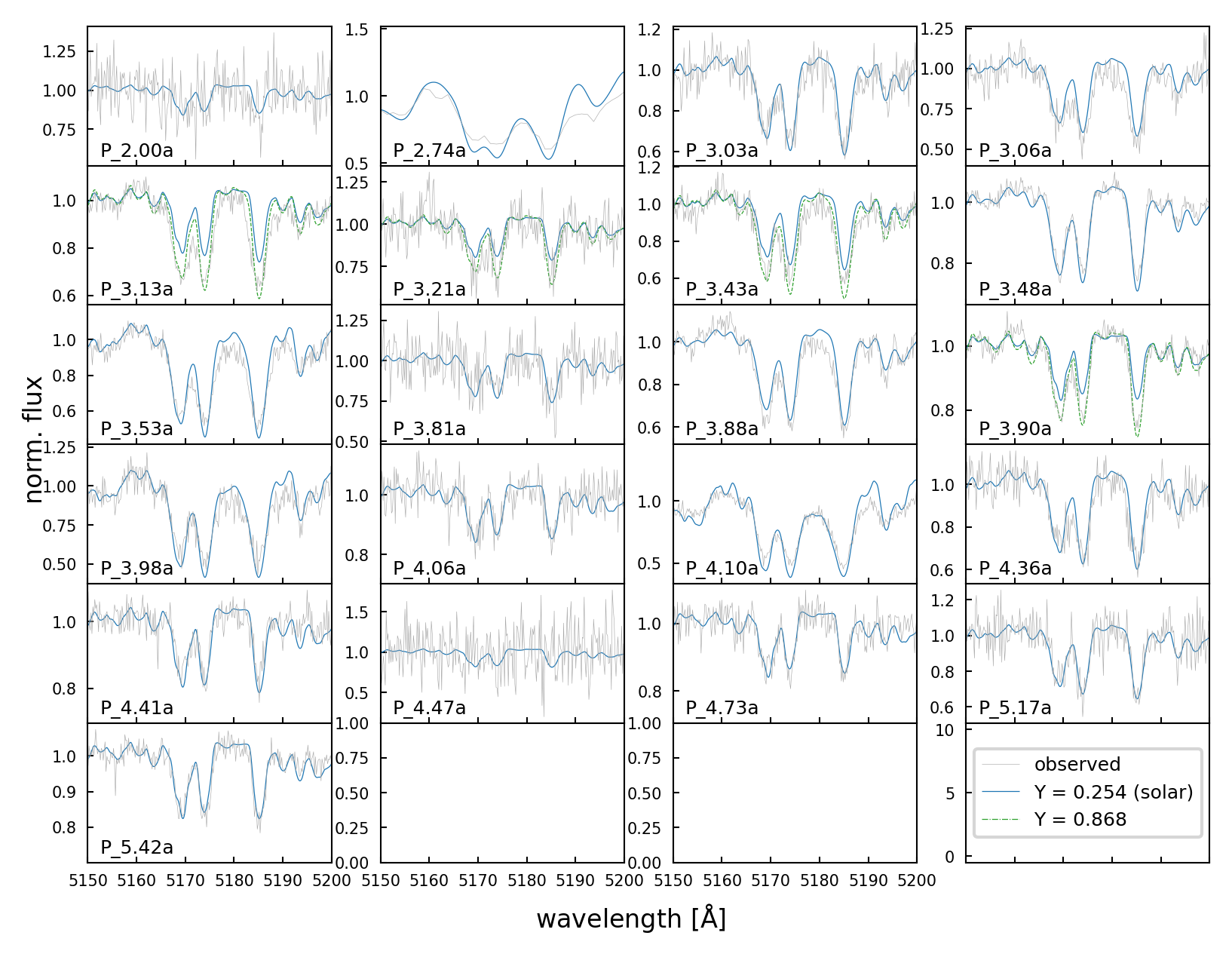}
    \caption{The Mg I triplet for all objects, plotted with their corresponding model spectra, used in the cross-correlation to obtain RVs. For four objects (P\_3.13a, P\_3.21a, P\_3.43a, P\_3.90a), the triplet is noticeably deeper than the standard solar abundance models and fits better to those with He mass fraction $Y = 0.868$. These are used later in the calculation of abundances.}
    \label{fig:mg_triplet}
\end{figure*}

For each spectrum, the radial velocity was obtained using the cross-correlation function (CCF) method \citep{Tonry1979AJ}, which we implemented as described in Appendix A of \citet{Zhang2021ApJS}. This involved shifting the spectra in log space of wavelength with a range of radial velocities corresponding to -600 and 600 km/s and finding the shift that maximized the CCF when compared to a standard stellar model spectrum. For measuring RVs, we used the wavelength range 4900-5400\AA, \ which includes the  MgI triplet (5167, 5173, 5184\AA), shown for all objects in Figure \ref{fig:mg_triplet}. For the model spectra, we used BOSZ Kurucz models \citep{Bohlin2017AJ} with solar abundances and log(g) and $T_{\rm eff}$ values close to those of the target. We retrieved models with $R=50,000$ and broadenend them to the resolution of each observed spectrum.  Instrumental broadening was carried out using convolution with a 1D Gaussian filter kernel \citep{AstropyCollaboration2022ApJ} with FWHM obtained from a gaussian fit to the [OI]5577\AA \ line in the sky spectra. 

We inferred projected rotation velocities $v\sin i$ for the donor in each system from the observed spectra (Section~\ref{sec:vsini}), but at this initial stage, we simply set $v\sin i = (2 \pi R/P_{\rm orb})\sin(i)$ using values from Table \ref{tab:BoE_info}. Rotational broadening was implemented using the \texttt{rotBroad} function \citep[as originally described by][]{Gray1992oasp} from PyAstronomy \citep{Czesla2019ascl},with linear limb-darkening coefficients taken from \citet{claret_gravity_2011}. RV errors were obtained by calculating chi-square ($\chi^2 = \sum \left(F_{\rm obs} - F_{\rm model}\right)^2/{\sigma_{\rm obs}^2}$) values between the observed and model spectra at each shift (for the wavelength range considered) and identifying the points at which $\chi^2$ increased by 1 from the minimum value. We note that while the radial velocity that minimizes $\chi^2$ is not precisely the same as that which maximizes the CCF, they typically only differ by $\lesssim 1$ km/s and the calculated errors should still provide a reasonable estimate of the statistical errors. 

Using the RV ephemerides from \citetalias{el-badry_birth_2021}, we can also calculate the expected RV of the donor at the orbital phase of our observed spectra. We compare these to the measured RVs in Table \ref{tab:RVs}. The typical disagreement between measured and predicted RVs is about 
30 km/s: small compared to the RV semi-amplitudes, but large compared to the formal uncertainties. We suspect that the main reason for disagreement between the measured and predicted RVs is that the RV zeropoint of the low-resolution spectra obtained by \citetalias{el-badry_birth_2021} was only stable at the $\sim 50$ km/s level (see their paper for details), leading to a systematic uncertainty in the binaries' center-of-mass RVs. The RV zeropoint for the ESI spectra analyzed in this work is stable at the few km/s level, so we expect our RVs to be more reliable than the predictions from \citetalias{el-badry_birth_2021}.

\begin{table}
    \centering
    \begin{tabular}{|c c c|}
         \hline
         ID & RV$_{\rm spec}$ [km/s] & RV$_{\rm phase, E21}$ [km/s] \\
         \hline
         P\_2.00a & -326.9$^{+5.2}_{-5.3}$ & -397.4 $\pm$ 10.8 \\
         P\_2.74a & -281.7$^{+3.3}_{-2.8}$ & -395.9 $\pm$ 15.6 \\
         P\_3.03a & -102.0$^{+2.6}_{-2.0}$ & -64.7 $\pm$ 10.4 \\
         P\_3.06a & 9.0$^{+1.4}_{-1.3}$ & 14.8 $\pm$ 9.0 \\
         P\_3.13a & -358.6$^{+5.7}_{-5.8}$ & -318.4 $\pm$ 18.4 \\
         P\_3.21a & -216.8$^{+1.8}_{-2.1}$ & -294.8 $\pm$ 12.5 \\
         P\_3.43a & -279.7$^{+2.1}_{-2.0}$ & -236.5 $\pm$ 9.2 \\
         P\_3.48a & 273.0$^{+1.8}_{-1.0}$ & 244.4 $\pm$ 5.4 \\
         P\_3.53a & 216.0$^{+8.1}_{-8.2}$ & 160.0 $\pm$ 29.2 \\
         P\_3.81a & -201.9$^{+2.4}_{-2.0}$ & -229.1 $\pm$ 5.1 \\
         P\_3.88a & 134.0$^{+1.3}_{-1.2}$ & 174.2 $\pm$ 8.0 \\
         P\_3.90a & -85.0$^{+1.8}_{-2.5}$ & -133.6 $\pm$ 3.4 \\
         P\_3.98a & 35.0$^{+3.9}_{-3.4}$ & -17.6 $\pm$ 8.3 \\
         P\_4.06a & -205.9$^{+0.7}_{-1.5}$ & -225.9 $\pm$ 8.6 \\
         P\_4.10a & 208.0$^{+4.1}_{-4.6}$ & 197.9 $\pm$ 24.9 \\
         P\_4.36a & 235.0$^{+2.5}_{-3.2}$ & 196.9 $\pm$ 7.5 \\
         P\_4.41a & -34.0$^{+2.6}_{-2.3}$ & -101.8 $\pm$ 3.0 \\
         P\_4.47a & 214.3$^{+30.3}_{-21.1}$ & 199.5 $\pm$ 17.5 \\
         P\_4.73a & -313.7$^{+1.9}_{-2.7}$ & -329.6 $\pm$ 6.7 \\
         P\_5.17a & -165.9$^{+1.1}_{-1.5}$ & -193.3 $\pm$ 11.0 \\
         P\_5.42a & 107.0$^{+4.1}_{-3.7}$ & 112.2 $\pm$ 6.3 \\
         \hline
    \end{tabular}
    \caption{Radial velocities of all objects. RV$_{\rm spec}$ are the results from CCF maximization using the spectra obtained in this work; RV$_{\rm phase, E21}$ are those calculated using the best fit orbital solutions from Table 3 of \citetalias{el-badry_birth_2021}. We find typical disagreements between the two RVs of $\sim 30$ km/s which can likely be attributed to the improved stability in the RV zeropoints of the ESI spectra compared to the lower resolution spectra from \citetalias{el-badry_birth_2021}.}
    \label{tab:RVs}
\end{table}

\subsection{Projected Rotational Velocity}
\label{sec:vsini}
As mentioned in \ref{ssec:rv}, we can get an expected value for the projected rotational velocity by using the values of radius, period, and inclination from \citetalias{el-badry_birth_2021}. However, this is only an estimate. In particular, the radii are obtained using fits to the spectral energy distributions (SEDs) while neglecting possible contamination from the companion or the accretion disk. Though comparisons to model spectra do not suggest significant contamination for most targets, strictly speaking, this makes the derived radii upper limits. 

\begin{figure}
    \centering
    \includegraphics[width=0.45\textwidth]{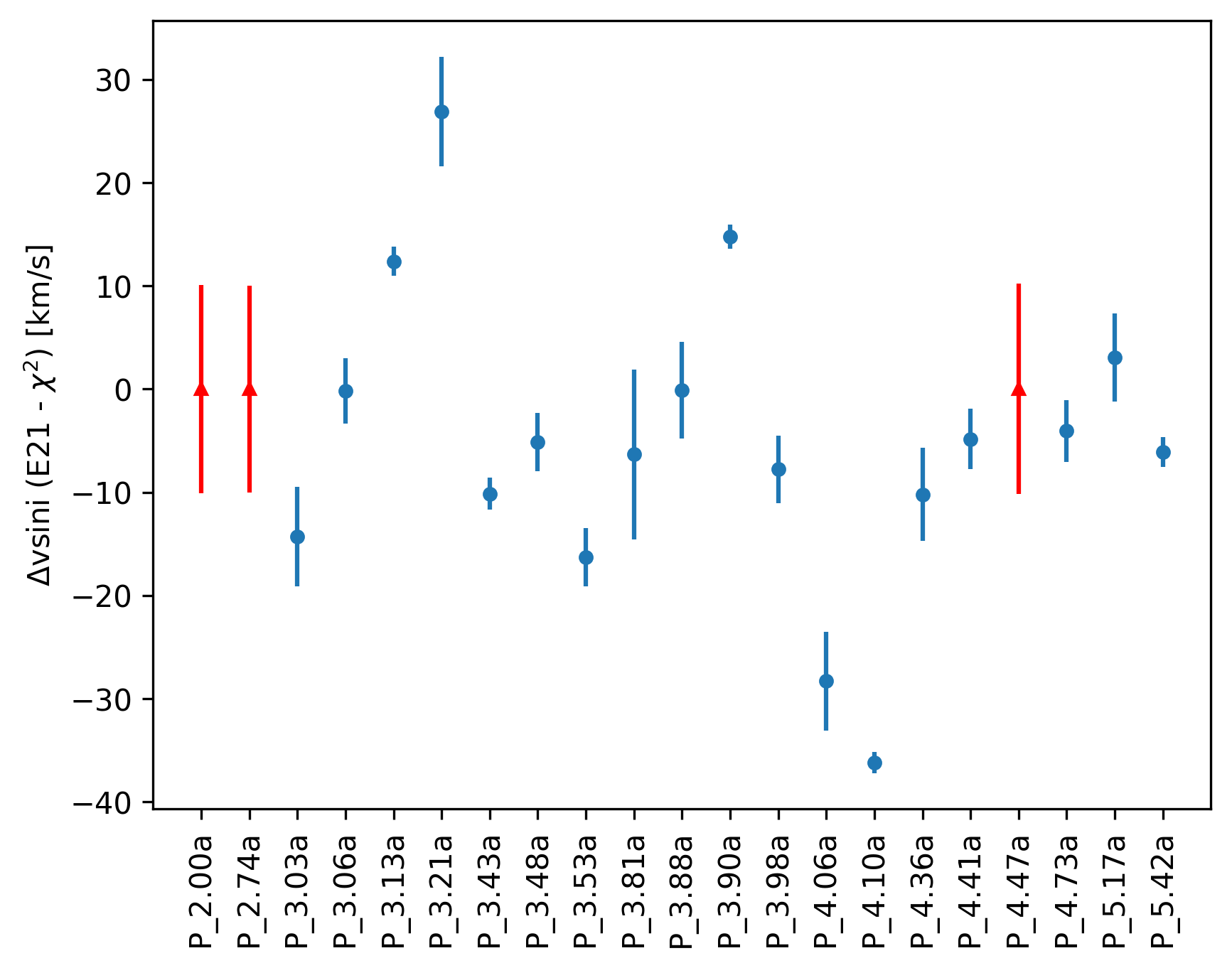}
    \caption{Difference in the $v \sin i$  calculated from $2 \pi R/P sin(i)$ using the values derived in \citetalias{el-badry_birth_2021} and that calculated with $\chi^2$ minimization. The points in red indicate objects for which no good solution was found using the $\chi^2$ method - they have been placed at 0 and their error bars have been arbitrarily extended to emphasize this. We see that most objects show an agreement of the two values to $\sim$ 30 km/s, suggesting that orbital parameters from \citetalias{el-badry_birth_2021} are reasonable estimates.}
    \label{fig:vsini_diff}
\end{figure}

Therefore, we also measured best-fit $v \sin i$ values directly from the rotational broadening of the observed spectra. For each target, this was done by first calculating radial velocities using model spectra rotationally broadened with a range of $v \sin i$ values from 50 to 150 km/s and shifting the target spectra with the derived RV. Then, the $\chi^2$ value was calculated by comparison with each model and the $v \sin i$ of the corresponding model that minimized the $\chi^2$ was identified. The errors were obtained by finding the location for which $\chi^2$ increased by one. Note that some of the parameters from \citetalias{el-badry_birth_2021} used to calculate the $v \sin i$ values have asymmetric errors. The errors on these have been approximated through standard error propagation using averages of the upper and lower values. The differences in the $v \sin i$ values from the two calculations are shown in Figure \ref{fig:vsini_diff}. 

We find that for three objects, the chi-square method did not result in clear local minima, instead monotonically decreasing towards larger $v \sin i$. These are marked in red in the figure, placed at the zero line (as the calculated differences have little meaning), and given arbitrarily large error bars for emphasis. This result is not unexpected given the particularly weak lines of P\_2.00a and P\_4.47a as well as the broad lines of P\_2.74a due to a lower resolution spectrum, all of which can be seen in Figure \ref{fig:mg_triplet}. Meanwhile, the rest of the objects have $v \sin i$ values that are in agreement to within about $\pm$ 10 km/s on average, suggesting that the radius and inclination constraints from from \citetalias{el-badry_birth_2021} are reliable.

\subsection{Equivalent Widths}

In order to measure the Na abundance in a donor, we first calculate the equivalent width (EW) of its absorption lines in the spectrum, in particular the 5900\AA \ doublet. Later, we repeat this process for many model spectra generated over a grid of abundances and make comparisons between the EWs of the models and the data (see Section \ref{sec:analysis}).

\begin{figure*}
    \centering
    \includegraphics[width=0.95\textwidth]{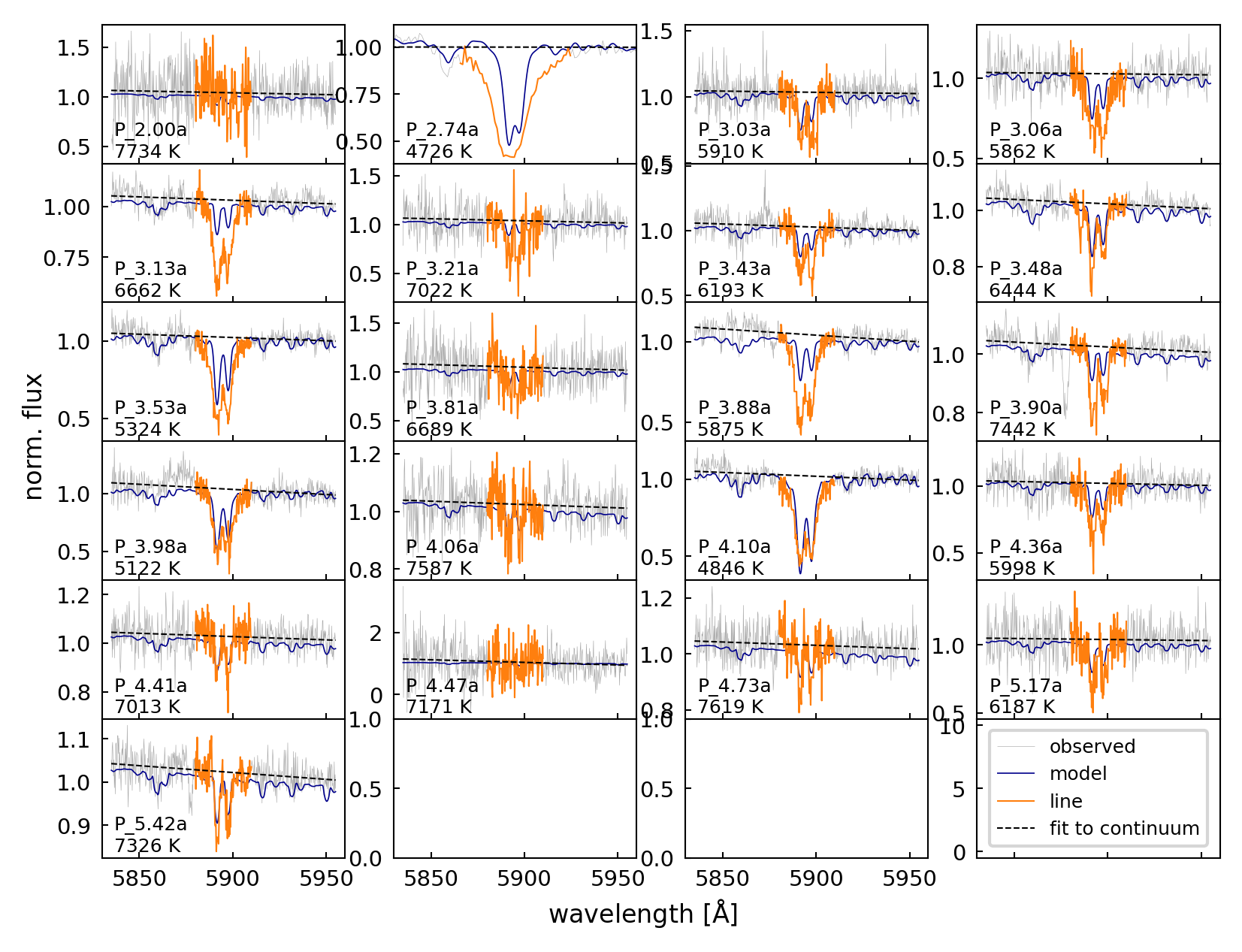}
    \caption{The Na 5900\AA \ doublet for all targets (for one of the two orders in which it is found). The region of the spectra over which the continuum is calculated is in gray, with the resulting first order polynomial fit in the black dashed line. The orange is the region considered to be occupied by the doublet and that under which the integrals, and thus EW, is calculated. A model spectra with $T_{\rm eff}$ matching each target (with [Na/H] = 0.0 dex) is also shown in blue for reference. The main takeaway from this figure is that the observed lines are noticeably deeper than those of the solar abundance model, indicating enhancement.}
    \label{fig:EW_regions}
\end{figure*}

To estimate the local continuum, the lines themselves were masked out, then a range 60\AA \ above and below the center of the doublet (120\AA \ for the one LRIS spectrum) was selected and fitted with a first order polynomial. The integral under the lines, $\pm$15\AA \ around the center ($\pm$30\AA \ for LRIS), was calculated numerically using the trapezoidal rule and subtracted from the area under the continuum over the same wavelength range to get the equivalent width (EW). Figure \ref{fig:EW_regions} shows these regions around the doublet for all of our objects. It is already clear by eye that for most targets, the observed line is significantly deeper and broader than that of the corresponding model spectrum with solar abundance. 

The error on the EW for each target was obtained by calculating the EWs of the same line in 100 Monte Carlo realizations of the observed spectra and taking their standard deviation. These were generated using the normalized flux of the original spectrum and its error as the standard deviation of a Gaussian distribution from which a random sample was drawn at each wavelength. 

\begin{figure*}
    \centering
    \includegraphics[width=\columnwidth]{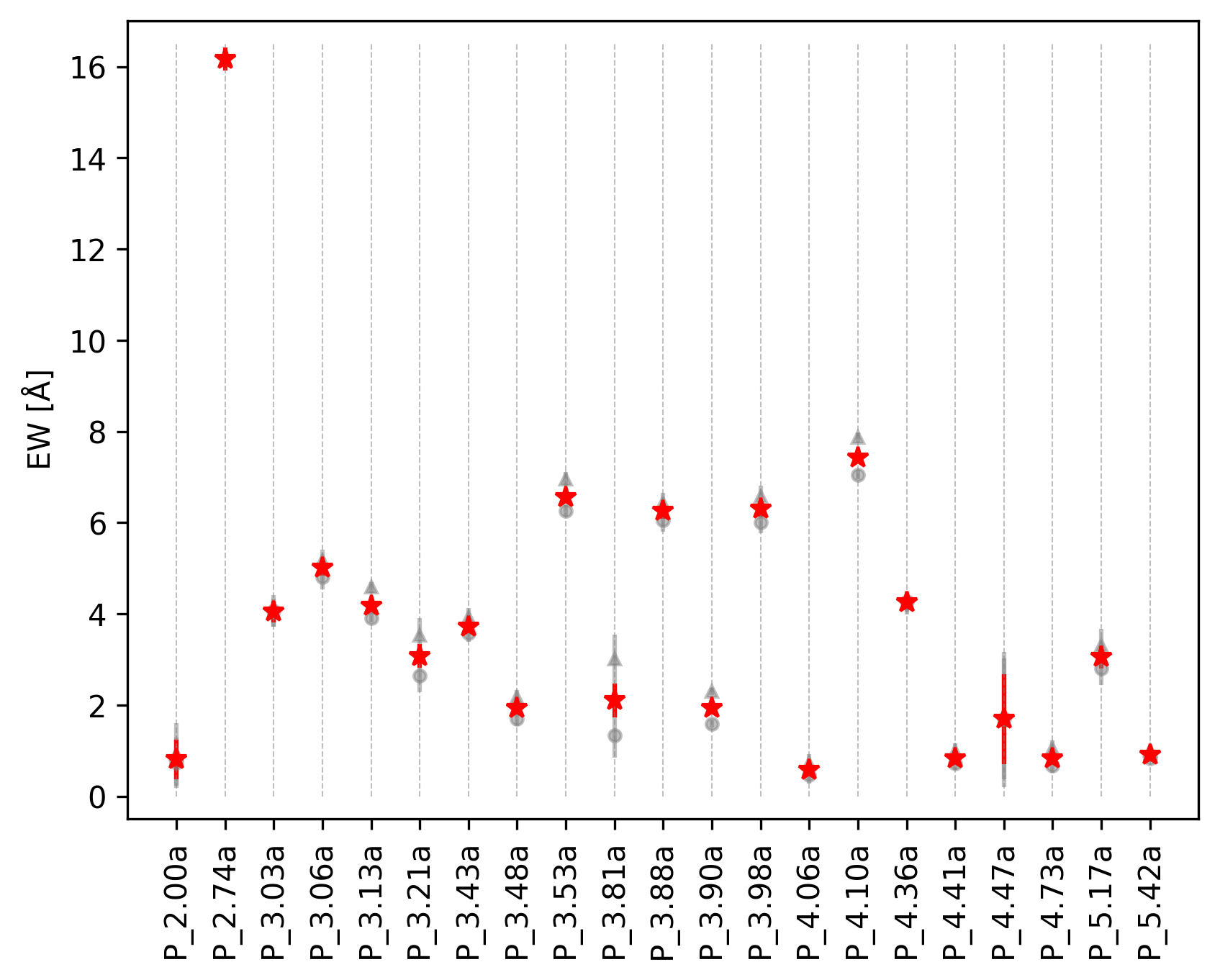}
    \includegraphics[width=\columnwidth]{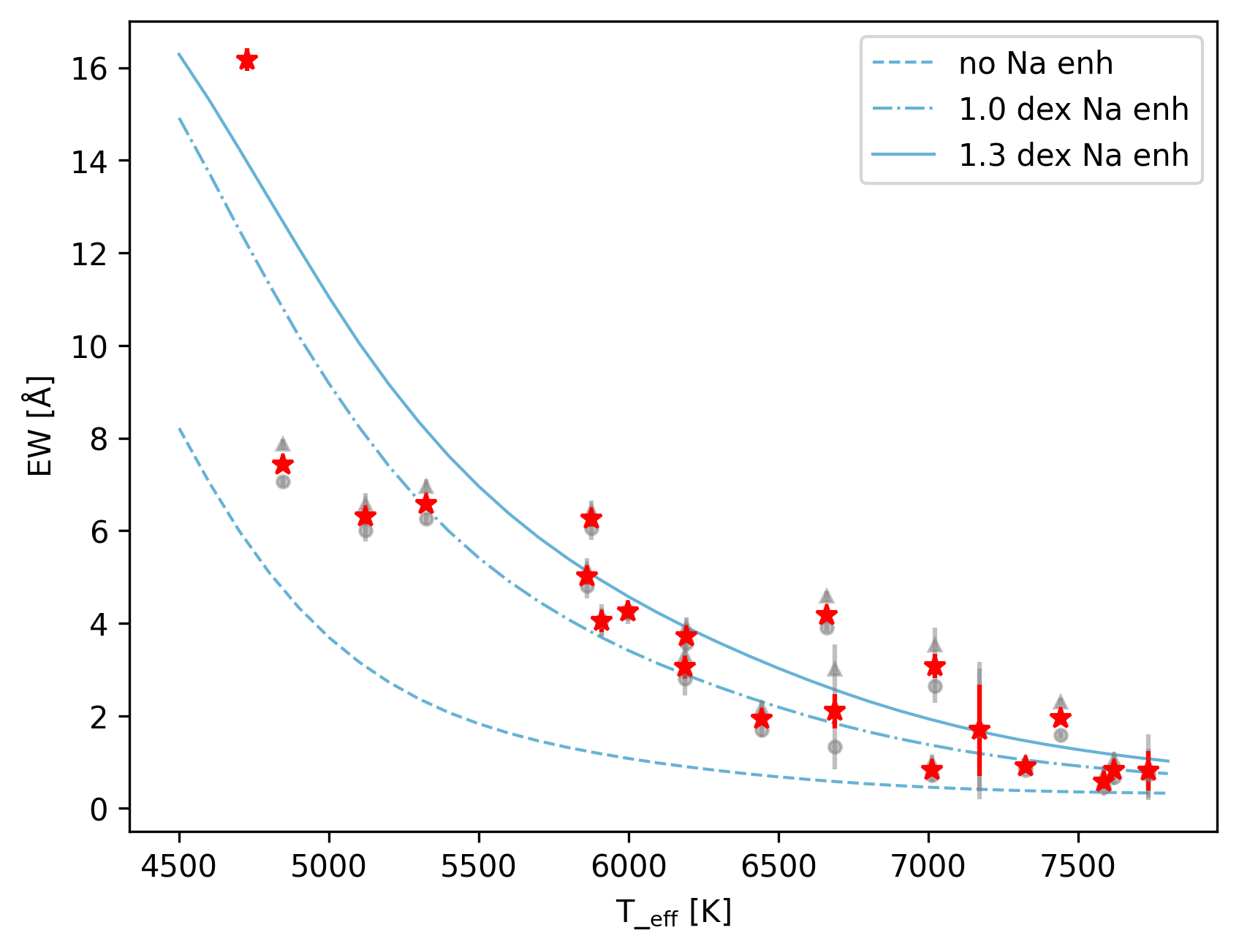}
    \caption{EWs of the Na 5900\AA \ doublet for all objects, plotted against their IDs (i.e. their orbital periods, \emph{left}) and their temperatures (\emph{right}). The gray points with the circle and triangle markers are the EWs of the line in each of the two orders while the red points with the star markers takes their weighted average. The solid, dash-dot, and dotted blues lines on the figure to the right plots the EW against temperature of model spectra with [Na/H] = 0.0, 1.0, and 1.3 dex respectively. All objects lie above the solar abundance line, meaning that they all show some level of enhancement, while more than half show $\gtrsim$ an order-of-magnitude enhancement.}
    \label{fig:EWs}
\end{figure*}

There are overlaps in the wavelength coverage towards the edge of each order for the ESI spectra. This means that some lines may be found in two orders, which is the case for the 5900\AA \ doublet. This allowed us to make the same calculations for the two orders and take their inverse-variance weighted average. For the one LRIS spectra (P\_2.74a), the EW was calculated just using the blue side where the line is located and no average was taken. Figure \ref{fig:EWs} plots the EWs of the 5900\AA \ doublet for all objects against their IDs and temperatures, for each of the two orders as well as their average. We see that there is a clear downward trend with increasing temperature: the lines become weaker at high $T_{\rm eff}$, where most of the Na is ionized. The EW values are also listed on Table \ref{tab:EWs_and_abun}, along with the derived Na abundances from comparison with models, described in Section \ref{ssec:abundances}.

\subsubsection{Accounting for interstellar or circumbinary absorption}

For many of the spectra, there were narrow and deep interstellar lines superimposed on top of the broad Na lines from the donors. Though they do not occupy a large area in most cases, to minimize their effect on our EW calculations, we cut out any data points $\pm$ 0.6\AA \ around the rest wavelengths of the doublet (as the interstellar lines have very small RV shifts) and replaced them with a median of several points above and below this range. This process effectively blocks out any obvious interstellar Na contribution which are present for a few objects (See Appendix \ref{appendix:IS}). This was not done for the LRIS spectrum because the resolution was too low that the doublet itself, and the interstellar lines, were not resolved and there were no distinguishable features on the single broad absorption line. 

In addition to absorption from the interstellar medium (ISM), it is in principle possible that there is additional Na absorption due to circumbinary material in the immediate vicinity of the targets. We explore this possibility -- and conclude that it is unlikely, because all the observed Na lines are broad and RV variable, tracking the donor -- in Appendix \ref{appendix:circumbinary_disk}. 

\section{Analysis} \label{sec:analysis}

\subsection{Kurucz model spectra} \label{ssec:kurucz_models}
To infer abundances from the observed spectra, we calculated a grid of 1D LTE model atmospheres and synthetic spectra for a range of effective temperatures and Na abundances, which we subsequently compared to the observed spectra. We used ATLAS 12 \citep{Kurucz1970SAOSR, kurucz_model_1979, Kurucz1992IAUS} to compute the atmosphere structure and SYNTHE \citep{Kurucz1993sssp} for the radiative transfer calculations, self-consistently re-computing the atmosphere structure for each model spectra. We use the linelist maintained by R. Kurucz\footnote{http://kurucz.harvard.edu/linelists.html} and assumed a microturbulent velocity of $2\,\rm km\,s^{-1}$. We generated spectra at resolution $R=300,000$ and applied instrumental and rotational broadening to match the observed data. 

We assumed a surface gravity $\log\left[g/\left({\rm cm\,s^{-2}}\right)\right]=4.8$ dex for all models, as the expected value for all the binaries in our sample is within 0.2 dex of this value. We generated models with a spacing of 100 K in $T_{\rm eff}$, ranging from 4500 to 7700 K, and a spacing of 0.1 dex in [Na/H], ranging from -0.1 to 1.3 dex. Our fiducial calculations assumed the solar abundance pattern for elements besides Na. As described below, we also generate several grid with He enrichment to explore its effect on the Na lines. 

\subsection{Helium enhanced models} \label{ssec:he_enh}

From Figure \ref{fig:mg_triplet}, we see that for four objects in particular - P\_3.13a, P\_3.21a, P\_3.43a, P\_3.90a - the observed Mg lines are noticeably deeper than those of the model with solar abundances. The same is true for other metal lines in these objects. 

One possible explanation for this is that the T$_{\rm eff}$ values inferred from SED fitting are overestimated. A higher T$_{\rm eff}$ means more of the Mg will be ionized, thus weakening its neutral lines. To explore this possibility, we compared the observed spectra with several models with lower T$_{\rm eff}$ and found that models cooler by $\approx 750$ K better matched the depth of the observed Mg triplets. However, these lower temperatures (with a corresponding increase in radius) resulted in a much worse fit to the broadband SEDs (Figure 9 in \citetalias{el-badry_birth_2021}), even when plausible light contributions from the accreting WD and/or disk were taken into account. We thus consider it unlikely that a too-cool assumed $T_{\rm eff}$ is the reason for the deeper metal lines in these objects. 

An alternative explanation for the deeper-than-expected Mg lines in these sources is surface helium enhancement of the donors. Evidence of enhancement has been observed in the donors of some other evolved CVs \citep[e.g.][]{harrison_identification_2018}, and some enhancement is also predicted by evolutionary models (Section~\ref{sec:mesa}). Lending support to the helium enhancement hypothesis, the same objects that have deeper-than-expected Mg lines also have deeper He I lines at 5876\AA\,\, than predicted by the fiducial spectral models (Figure~\ref{fig:EW_regions}).  

To analyze these objects and assess the sensitivity of Na abundance measurements to the assumed helium abundance, we generated several grids of models (with the same spacing in $[Na/H]$ and $T_{\rm eff}$ as before) with surface He mass fractions, $Y$, of 0.569 and 0.868 (compared to the solar value of 0.254). We plot the models with $Y = 0.868$ in Figure \ref{fig:mg_triplet} for the four objects and we see that these have deeper Mg I triplet lines than the model with $Y = 0.254$, in better agreement with the observed spectra. Thus, we carry out the calculation of Na abundances for these objects in Section \ref{ssec:abundances} using these He enhanced models. 

For the remaining objects, the good agreement of the observed spectra and models with solar helium abundances rules out strong helium enhancement, so we use the solar models. However, it is possible that some modest enhancement could go unrecognized. We explore how unaccounted-for He enhancement would affect our inferred Na abundances in Appendix \ref{appendix:He_enh}, where we conclude that the uncertainty in our inferred Na abundances associated with possible He enhancement is typically about 0.1 dex. 

\subsection{Abundance calculations} \label{ssec:abundances}

\begin{figure}
    \centering
    \includegraphics[width=\columnwidth]{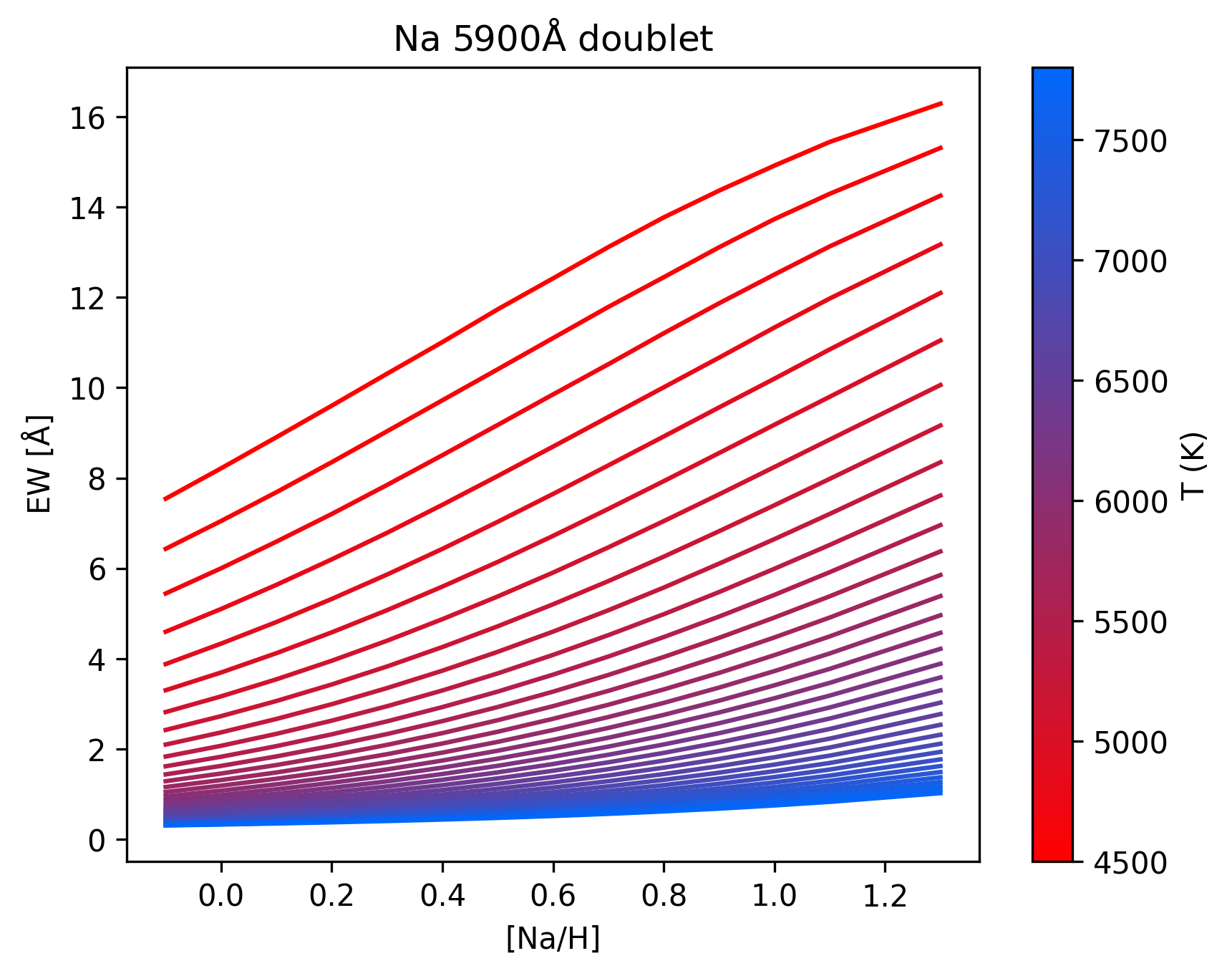}
    \caption{EWs of the Na 5900\AA \ doublet against the Na abundances for model spectra with a range of effective temperatures. At a given $T_{\rm eff}$, we see that EW increases monotonically with [Na/H]. The slope gets shallower at higher $T_{\rm eff}$ as more Na gets ionized.}
    \label{fig:abun_models}
\end{figure}

We calculate the EWs of the same Na lines for all model spectra with different abundances at each $T_{\rm eff}$. This is plotted on Figure \ref{fig:abun_models}. At all $T_{\rm eff}$, we find that with more Na (i.e. higher [Na/H]), the EW of its line is greater. However, the sensitivity of this relation depends strongly on the $T_{\rm eff}$ whereby the overall slope gets less steep at higher temperatures as more of the Na is ionized. 

For each object, we derive the relation between EW and abundance at its temperature $T_{\rm eff,0}$ using interpolation between the relations for two models lying above and below $T_{\rm eff,0}$. The corresponding abundance is then obtained at the EW of the line measured from its spectra. The error is calculated by propagating the errors in the EW and $T_{\rm eff}$. We note that the uncertainty in $T_{\rm eff}$ obtained from SED fitting (listed on Table 2 of \citetalias{el-badry_birth_2021}, ranging from $\sim 50 - 100$ K) is likely underestimated due to various systematics so we instead take it to be $100$ K for all objects. 

\begin{figure*}
    \centering
    \includegraphics[width=0.95\textwidth]{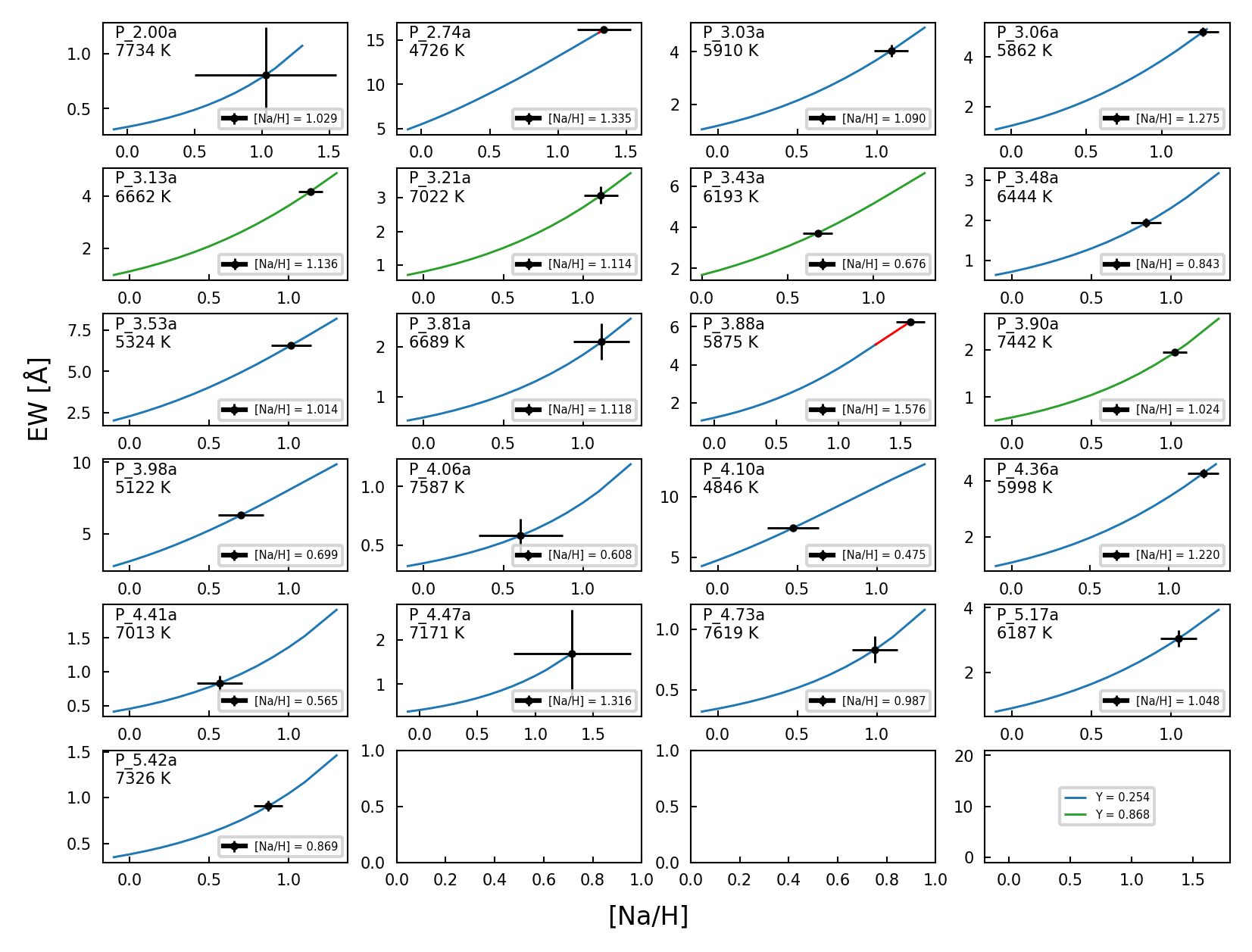}
    \caption{Plots of EW against [Na/H] obtained by interpolating those in Figure \ref{fig:abun_models} to the temperature of each object. The red lines indicate extrapolation, where the calculated EW of the object is above the maximum value that was generated. The blue and green lines corresponds to models with solar and enhanced He mass fractions. Most objects have [Na/H] ranging from $\sim$ 0.4 - 1.3 dex, with the exception of P\_3.88 with a highly extrapolated value of 1.576 dex.}
    \label{fig:Na_abun}
\end{figure*}

The resulting plots are shown for all objects in Figure \ref{fig:Na_abun}. Green plots indicate that that the objects are being compared to He enhanced models with $Y = 0.868$, as discussed in Section \ref{ssec:he_enh}.
The abundances with their errors are also summarized in Table \ref{tab:EWs_and_abun}. We see that all objects show some level of Na enhancement with [Na/H] values ranging from $\sim$ 0.4 - 1.6 dex, with a median value at a significant 1.024 dex i.e. about an order of magnitude enhancement compared to solar value.

\subsection{Non-LTE corrections} \label{ssec:NLTE}

In generating the curves of growth (Figures \ref{fig:abun_models} and \ref{fig:Na_abun}) used to calculate the Na abundances of our objects, we neglected deviations from the standard assumption of local thermodynamic equilibrium (LTE) used in the model atmospheres. It has been found that the inclusion of non-LTE (NLTE) effects can increase the EWs of Na lines at fixed [Na/H] \citep[e.g.][]{Mashonkina2000ARep, Lind2011A&A, Lind2022A&A}. To avoid overestimating the abundances, we apply NTLE corrections to our abundances using the results from \citet{Lind2022A&A}. 

\begin{figure}
    \centering
    \includegraphics[width=0.45\textwidth]{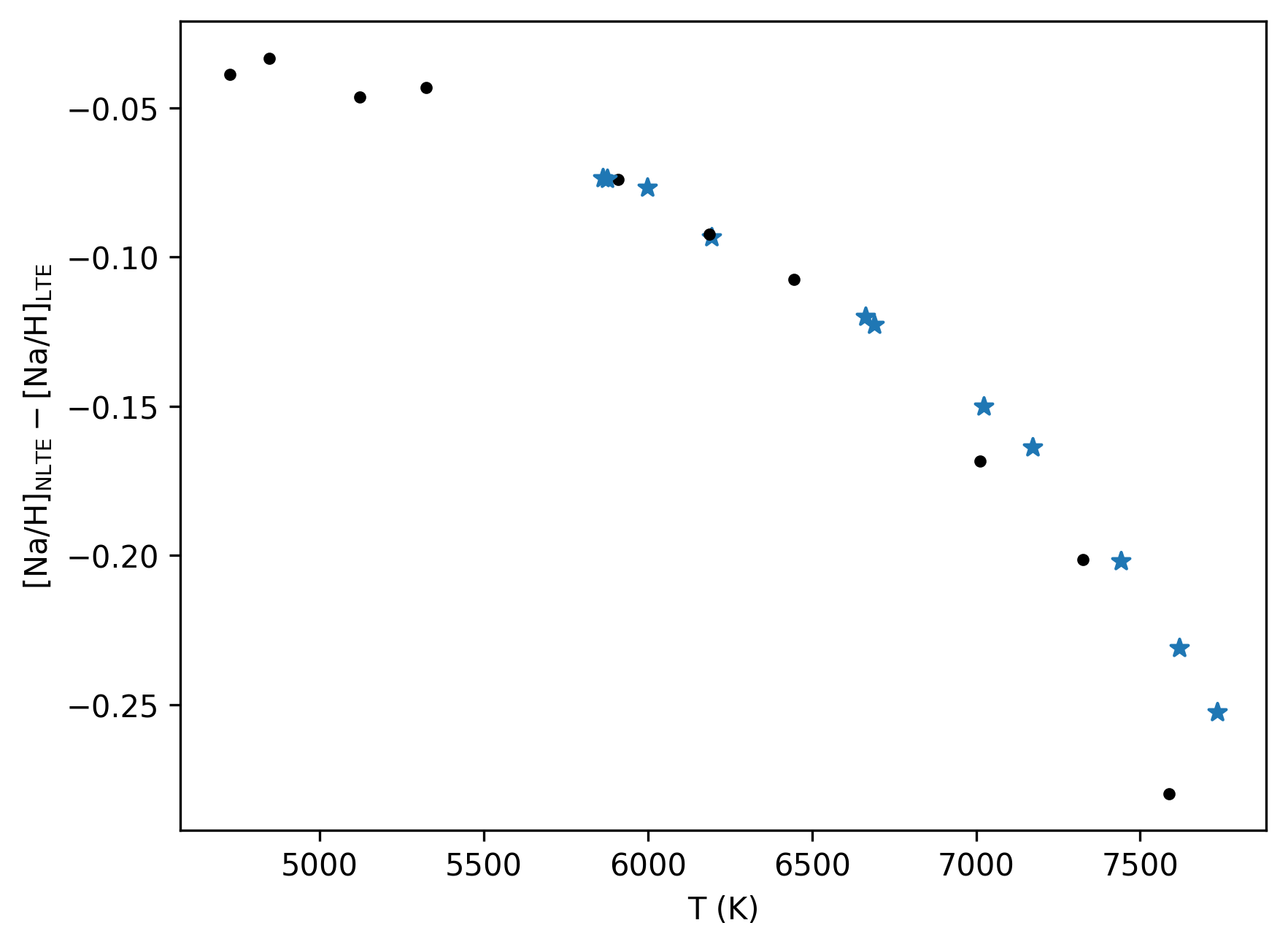}
    \caption{The difference in the Na abundances implied from EWs of the Na 5900 \AA \ doublet of synthetic spectra with and without NLTE effects, as a function of effective temperature for all objects. The blue stars indicate objects whose EWs were greater than the maximum value provided by \citet{Lind2022A&A} for which the difference plotted is that at the maximum value. In all cases, we see that inclusion of NTLE effects leads to a reduction in the [Na/H], with a mean value of -0.126 dex, so neglecting them will result in an overestimate of the Na enhancements.}
    \label{fig:NLTE}
\end{figure}

At the measured EWs of all objects (sum of the 5890 and 5896 \AA \ lines), we calculate the difference between the predicted NLTE and LTE abundances, interpolated to their effective temperatures. We used values for [Fe/H] = 0.0 dex, log(g) = 5.0, and microturbulence of 2 km/s. The thus-inferred NLTE corrections are shown as a function of temperature in Figure \ref{fig:NLTE}. The mean correction is $-0.126$ dex, and the correction is more important at higher temperatures. We add these corrections to our previously measured abundances in Section \ref{ssec:abundances}. These ``NLTE-corrected" abundances can be found on Table \ref{tab:EWs_and_abun}. These values range from $\sim$ 0.3 - 1.5 dex, with a median value of 0.956 dex, and will be used when referring to [Na/H] in the rest of the paper. 

We emphasize that these are approximate corrections, done to avoid overstating the Na enhancements. The LTE curves of growth from \citet{Lind2022A&A} are similar but not identical to those shown in Figure \ref{fig:abun_models} as the synthetic spectra were not calculated with the same code. For several objects, indicated with blue stars on Figure \ref{fig:NLTE}, our measured EWs exceeded the maximum value calculated by \citet{Lind2022A&A} (which corresponded to [Na/H]=0.7) so we simply used the NLTE correction at the maximum value. However, as the absolute differences get smaller towards larger EWs, this only results in more conservative values for the final abundances. Lastly, it should be noted that as discussed in Section \ref{ssec:he_enh}, some of our objects were modelled with He enhancement but this was not taken into account in calculating the NLTE corrections.

\begin{table}
    \centering
    \begin{tabular}{|c c c c|}
         \hline
         ID & EW [\AA] & [Na/H] & [Na/H]$_{\rm NLTE}$ \\
         \hline
         P\_2.00a & 0.808 $\pm$ 0.429 & 1.029 $\pm$ 0.526 & 0.777 $\pm$ 0.526 \\
         P\_2.74a & 16.176 $\pm$ 0.251 & 1.335 $\pm$ 0.197 & 1.296 $\pm$ 0.197 \\
         P\_3.03a & 4.043 $\pm$ 0.231 & 1.090 $\pm$ 0.105 & 1.016 $\pm$ 0.105 \\
         P\_3.06a & 5.014 $\pm$ 0.174 & 1.275 $\pm$ 0.104 & 1.201 $\pm$ 0.104 \\
         P\_3.13a & 4.177 $\pm$ 0.078 & 1.136 $\pm$ 0.076 * & 1.016 $\pm$ 0.076 * \\
         P\_3.21a & 3.074 $\pm$ 0.258 & 1.114 $\pm$ 0.108 * & 0.964 $\pm$ 0.108 * \\
         P\_3.43a & 3.723 $\pm$ 0.138 & 0.676 $\pm$ 0.085 * & 0.583 $\pm$ 0.085 * \\
         P\_3.48a & 1.936 $\pm$ 0.111 & 0.843 $\pm$ 0.096 & 0.736 $\pm$ 0.096 \\
         P\_3.53a & 6.566 $\pm$ 0.098 & 1.014 $\pm$ 0.127 & 0.970 $\pm$ 0.127 \\
         P\_3.81a & 2.102 $\pm$ 0.366 & 1.118 $\pm$ 0.177 & 0.995 $\pm$ 0.177 \\
         P\_3.88a & 6.251 $\pm$ 0.166 & 1.576 $\pm$ 0.116 & 1.502 $\pm$ 0.116 \\
         P\_3.90a & 1.941 $\pm$ 0.055 & 1.024 $\pm$ 0.078 * & 0.822 $\pm$ 0.078 * \\
         P\_3.98a & 6.298 $\pm$ 0.168 & 0.699 $\pm$ 0.142 & 0.653 $\pm$ 0.142 \\
         P\_4.06a & 0.581 $\pm$ 0.142 & 0.608 $\pm$ 0.265 & 0.328 $\pm$ 0.265 \\
         P\_4.10a & 7.424 $\pm$ 0.073 & 0.475 $\pm$ 0.162 & 0.442 $\pm$ 0.162 \\
         P\_4.36a & 4.260 $\pm$ 0.169 & 1.220 $\pm$ 0.098 & 1.143 $\pm$ 0.098 \\
         P\_4.41a & 0.834 $\pm$ 0.111 & 0.565 $\pm$ 0.141 & 0.397 $\pm$ 0.141 \\
         P\_4.47a & 1.692 $\pm$ 0.990 & 1.316 $\pm$ 0.506 & 1.152 $\pm$ 0.506 \\
         P\_4.73a & 0.833 $\pm$ 0.110 & 0.987 $\pm$ 0.143 & 0.756 $\pm$ 0.143 \\
         P\_5.17a & 3.054 $\pm$ 0.259 & 1.048 $\pm$ 0.115 & 0.956 $\pm$ 0.115 \\
         P\_5.42a & 0.908 $\pm$ 0.057 & 0.869 $\pm$ 0.091 & 0.668 $\pm$ 0.091 \\
         \hline
    \end{tabular}
    \caption{Table of the EWs of the 5900\AA \ doublet with the derived [Na/H] from Figure \ref{fig:Na_abun}. The [Na/H]$_{\rm NLTE}$ has been corrected for NTLE effects using values from Figure \ref{fig:NLTE}, calculated using curves of growth provided in \citet{Lind2022A&A}. [Na/H] values were calculated using solar He abundance (Y = 0.254) models, except those marked with an asterisk (*) which were calculated using He enhanced (Y = 0.868) models. [Na/H] of objects range from $\sim$ 0.4 - 1.6 dex, with a median value of 1.026 dex. Meanwhile, [Na/H]$_{\rm NLTE}$ range from $\sim$ 0.3 - 1.5 dex, with the median at 0.956 dex.}
    \label{tab:EWs_and_abun}
\end{table}

\section{MESA evolutionary models} \label{sec:mesa}

We ran MESA models \citep{Paxton2011ApJS, Paxton2013ApJS, Paxton2015ApJS, Paxton2018ApJS, Paxton2019ApJS} to study the evolutionary history of the evolved CVs and to test whether they predict the observed overabundance of Na. Similar to the models used in \citetalias{el-badry_birth_2021} (described in greater detail in \citealt{el-badry_lamost_2021} and summarized here), we only follow the evolution of the donor/secondary and model the WD companion/primary as a point mass. For the mass transfer rates, we use those of optically thick Roche lobe overflow described by \citet{Kolb1990A&A}. We consider fully non-conservative mass transfer meaning that all of the mass transferred to the primary is instantaneously ejected from its vicinity as fast winds, described by the $\beta$ parameter being set to 1 (note that MESA uses $\beta$ as defined by \citealt{Tauris2006csxs} which is 1-$\beta$ as defined earlier by \citealt{rappaport_evolution_1982}). This is a reasonable assumption in the case of CVs where WDs periodically lose accreted mass by classical novae explosions on timescales short to those over which the orbit evolves. We include the torque applied by magnetic braking as described by equation 36 of \citet{rappaport_new_1983} and set the gamma exponent to 3. The strength of magnetic braking in CVs at the periods of our sample are uncertain e.g. \citep[e.g.][]{el-badry_magnetic_2022}, but the adopted magnetic braking law primarily affects how quickly CVs evolve, not their abundances at a fixed evolutionary state. 

We made several modifications compared to the MESA models described by \citetalias{el-badry_birth_2021}. Firstly, we used a newer version of MESA (r22.05.1) which called for changes to the options in the `star' module of the secondary. To initialize it as a ZAMS star with solar abundances from \citet{grevesse_no_1998}, we used the set\_uniform\_initial\_composition option and manually defined the initial mass fractions of H, He, and metals (0.70, 0.28, 0.02 respectively). We also turned on rotation with the surface velocity set by tidal synchronization (which applies at the short orbital periods of our objects). 
Most importantly, MESA’s default nuclear reaction network (\texttt{basic.net}) does not include Na and thus we instead use the network \texttt{sagb\_NeNa\_MgAl.net} which follows 22 isotopes from $^1$H to $^{27}$Al, including $^{21-23}$Na. It uses rates provided by the JINA reaclib database V2.0 \citep{farmer_carbon_2015}. 
The transitions between the CNO, NeNa, and MgAl cycles are depicted in Figure 1 of \citet{boeltzig_shell_2016}. The step producing $^{23}$Na is proton capture by $^{22}$Ne or in shorthand notation, $^{22}$Ne(p,$\gamma$)$^{23}$Na. The NeNa and MgAl cycles are often discussed in the case of AGB stars undergoing Hot Bottom Burning (HBB) where they are activated at the high temperatures reached at the base of the convective layer \citep{izzard_reaction_2007} as well as in novae outbursts \citep{jose_nuclear_1999}, but $^{23}$Na is generically produced as long as sufficiently high temperatures are reached. If the NeNa cycle produces excess Na in the cores of CV donors during their main-sequence evolution, the resulting Na enhancement can be observed on the stars' surfaces after  their outer layers are stripped.

We calculated models with initial donor masses of $1.0\,M_{\odot}$ and $1.5\,M_{\odot}$, assuming a WD mass $M_{\rm WD}=0.7\,M_{\odot}$ in all cases. 
For each $M_{\rm donor}$, we ran models over a grid of initial periods, $p_0$ (ranging from $\sim 0.5 - 5$ days), to find the initial periods which would produce systems with properties close to those of the evolved CVs in our sample. We did not consider lower initial donor masses here because such donors would take more than a Hubble time to complete their main-sequence evolution and become evolved CVs. Meanwhile, donors with higher initial masses would undergo a phase of thermal-timescale mass transfer shortly after initial Roche lobe overflow, resulting in a high mass transfer rate and stable H burning on the surface of the accreting WD \citep[e.g.][]{Schenker2002}. This would lead to an increase in the WD's mass that would not be reliably followed by our calculations, which assume mass transfer is fully non-conservative on long timescales. 

\begin{figure*}
    \centering
    \includegraphics[width=0.9\textwidth]{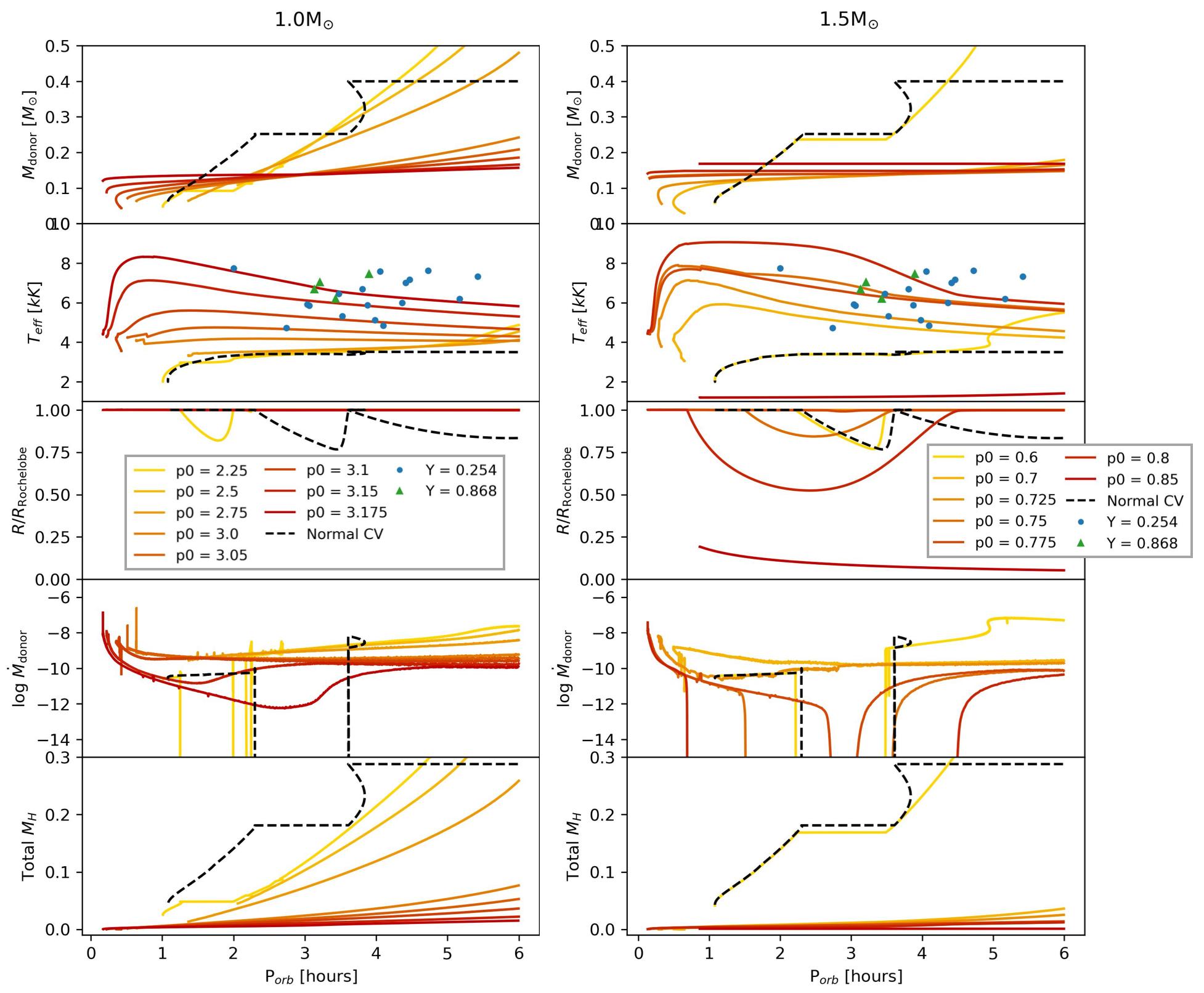}
    \caption{MESA calculations for initial donor masses of $1.0 \ M_{\odot}$ (Left) and $1.5 \ M_{\odot}$ (Right), for a range of initial periods, $p_0$. The figures plot the mass, effective temperature, Roche-Lobe filling factor, mass transfer rate, and remaining hydrogen mass of the donor/secondary star against the orbital period. We also add the tracks for a normal CV with an initial donor mass of $0.4 \ M_{\odot}$ and $p_0 = 0.3$ days (black dashed line). The effective temperatures of our objects are also plotted over the tracks for reference (blue and green points). We see that these temperatures are significantly higher than that of a normal CV. For the $1.0 M_{\odot}$ model, the $p_0$ range which result in the temperatures of our objects ($\sim 4500-7500$ K) at their orbital periods ($\sim 2-6$ h) is $\sim 3.0 - 3.2$ days. For the $1.5 \ M_{\odot}$ model, this range is $\sim 0.7 - 0.8$ days.}
    \label{fig:physical}
\end{figure*}

\begin{figure*}
    \centering
    \includegraphics[width=0.85\textwidth]{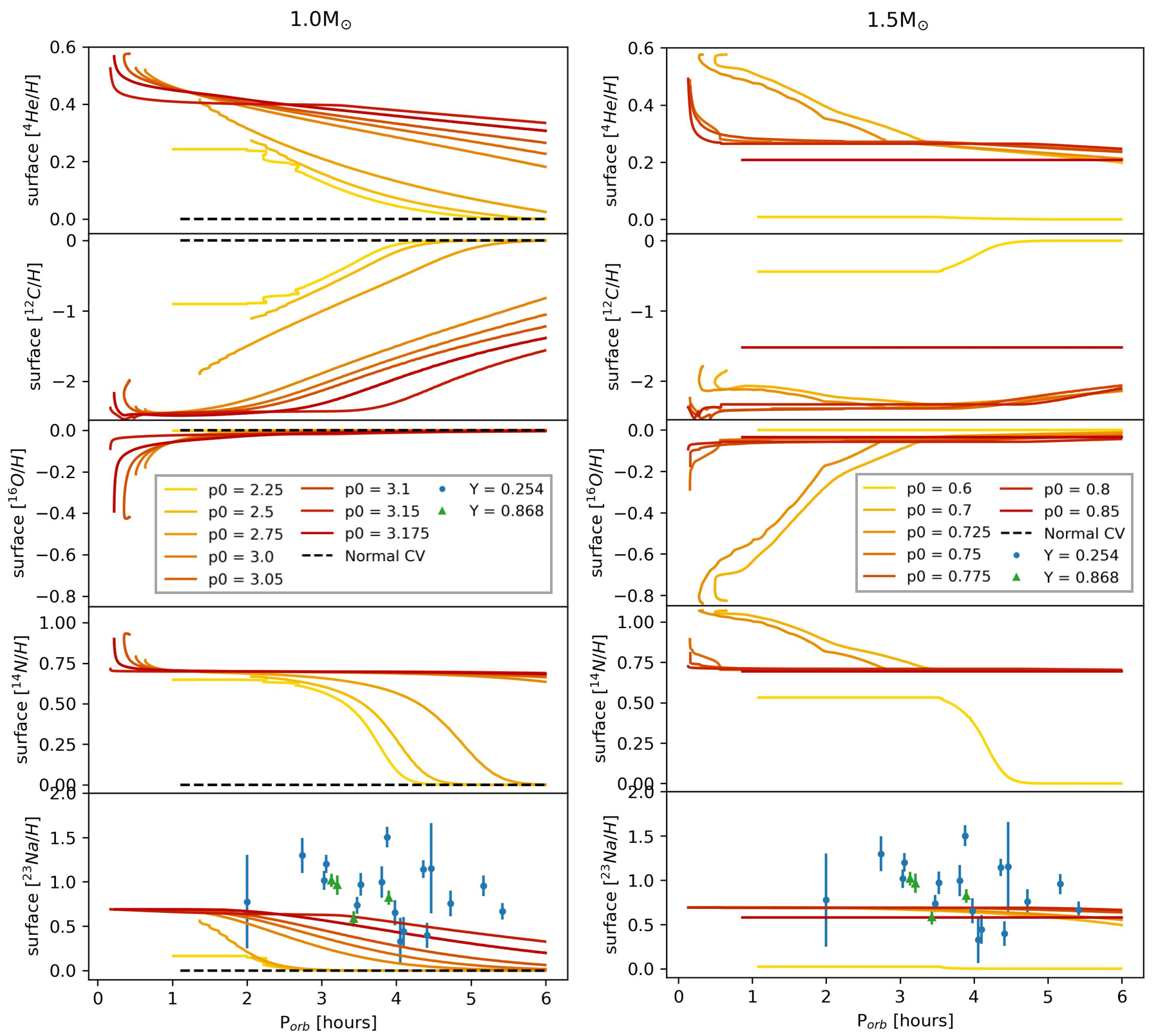}
    \caption{Surface abundances of $^{4}$He, $^{12}$C, $^{16}$O, $^{14}$N, and $^{23}$Na against orbital period for the MESA models in Figure \ref{fig:physical}. The points in the plots of $^{23}$Na abundance are those of the objects from this paper ([Na/H]$_{\rm NLTE}$ from Table \ref{tab:EWs_and_abun}) to serve as comparison to those predicted by the models (the blue and green points indicate abundances calculating using solar He abundance (Y = 0.254) and He enhanced (Y = 0.868) models, respectively). It is seen that while the MESA models do predict significant Na enhancements, they generally underpredict them compared to observed values.}
    \label{fig:chemical}
\end{figure*}

Figure \ref{fig:physical} shows the evolution of several basic parameters of the donor - the mass, effective temperature, Roche lobe filling factor, mass loss rate, and total hydrogen mass - as a function of the orbital period for a range of initial periods. For reference, we also add tracks for a "normal" (i.e. not evolved) CV (black dashed lines) with initial with $M_{\rm donor} = 0.4 \ M_{\odot}$ and $p_0 = 0.3$ days. With increasing $p_0$, the donor takes longer to become Roche lobe filling so that it is more evolved at the onset of mass transfer (although this figure is restricted to the short orbital periods of our objects where most models are already mass-transferring i.e. R/R$_{\rm Roche Lobe} \sim $ 1 and log $ \dot{M}_{\rm donor} < $ 0, plotted on the third and fourth panels respectively). Therefore, systems with a longer $p_0$ end up reaching higher effective temperatures at short periods $\lesssim 5$ h, as seen on the second panels. The last panels show that they are also left with smaller total masses of remaining hydrogen as they have undergone more H core burning while on the main sequence. 

For initial $M_{\rm donor} = 1.0 \ M_{\odot}$, models with  $p_0 \gtrsim 2.5$ days remain mass transferring across the period gap ($P_{\rm orb} \approx 2-3$ hours, where the normal CV (black, dashed) detach due to the donor becoming fully convective and magnetic braking weakening). This reflects the fact that these donors are sufficiently evolved for their helium-enriched cores to remain radiative so that they do not experience a disruption of magnetic braking at $P_{\rm orb} \approx 3$ hours and thus have continuous mass transfer. These objects also reach very short periods $\lesssim 1$ h where they would be observed as AM CVn binaries (e.g. \citealt{podsiadlowski_cataclysmic_2003}). The model with $p_0 = 2.25$ days has a somewhat less evolved donor that does detach from its Roche lobe when it becomes fully convective, but this only occurs at $P_{\rm orb} \sim 2$ h -- shorter than the 3 hour period where normal CVs with fully unevolved donors detach. We see that while a larger range of initial periods remain in contact across the gap, only the models with $3.0 \lesssim p_0 \lesssim 3.2$ days reach the range of effective temperatures at the observed periods of our sample, also plotted on the figure ($T_{\rm eff}\sim4500 - 7500$\,K at $P_{\rm orb} \sim 2-6$). Beyond $p_0 \approx 3.2$ days, the donor leaves the main sequence before mass transfer starts (described further below for the $1.5 \ M_{\odot}$, $p_0 = 0.85$ case). 

For initial $M_{\rm donor} = 1.5 \ M_{\odot}$, the range of initial periods forming evolved CVs that reach the temperatures of our objects is shorter and narrower, from $\sim 0.7 - 0.8$ days. This is expected because more massive stars are hotter and have thinner convective envelopes. This means that magnetic braking is weaker so the stars must be initially closer together to begin mass transfer as the donors leave the main sequence. Furthermore, they have shorter lifetimes and evolve more quickly off the main sequence. The $p_0 = 0.6$ model is a case where the donor is hardly evolved and thus its evolution closely follows that of a normal CV. Meanwhile, the $p_0 = 0.85$ model represents a system that is above the``bifurcation limit'' where the donor has had enough time for it to ascend the giant branch by the time it begins mass transfer \citep{podsiadlowski_cataclysmic_2003}. By the time the system reaches the period range of our objects, the donor has evolved into a WD and is on the cooling track, with gravitational radiation slowing bringing it closer to its companion until they are brought back in contact at very short periods. 
\citetalias{el-badry_birth_2021} also tested the effect of altering other initial parameters including the metallicity and the exponential overshooting parameter but found that, given the same degree of donor evolution at the onset of mass transfer, these parameters only modestly change the mass of the donor and qualitative evolutionary tracks at short periods. 

Figure \ref{fig:chemical} shows surface abundances of the donor for several elements against the orbital period. Once again, we add the plots for a normal CV in all panels which remain approximately constant at 0.0 dex (i.e. solar) for all elements. For the [Na/H] panel, we also plot points for the measured abundances of our objects on top of the models.

We see that $^{4}$He, $^{14}$N, and $^{23}$Na are enhanced while $^{12}$C and $^{16}$O are depleted at the observed periods. The He enhancement can be explained simply by the fact that we are seeing the core of an evolved donor that has undergone significant H burning and whose envelope has been stripped away as a result of the mass transfer. This also supports the possibility that enhanced He may be what is responsible in deepening of the magnesium lines (Section \ref{ssec:he_enh}; we note that $^{24}$Mg is one stable element in the MgAl cycle - while its abundance was traced in several models, there were no significant changes from the initial solar values as a result of this process). The abundances of C, N, and O can be explained by looking at the reactions in the CNO cycle of which $^{14}$N(p,$\gamma$)$^{15}$O is the slowest. This results in a build up of $^{14}$N and a corresponding exhaustion of C (and O to a lesser extent). As described earlier, the Na enhancement is likely the result of advanced H burning which only occur in the high temperatures of evolved stars. From the plots of [Na/H], we see that the abundance at $P_{\rm orb} \sim 2-6$ h is sensitive to the initial period. Thus, for a given initial donor mass, we are only able to obtain a maximum predicted value of [Na/H] (corresponding to the longest $p_0$ i.e. the most evolved model) at a specified orbital period. For $M_{\rm donor} = 1 \ M_{\odot}$, this is at [Na/H] $\sim 0.6$ dex and for $1.5 \ M_{\odot}$, this is greater at [Na/H] $\sim 0.7$ dex at $P_{\rm orb} \sim 4$ h. Thus, a system with an initially more massive secondary (within the mass range that leads to stable mass transfer) results in a greater Na enhancement at the observed periods as it reaches higher temperatures and thus undergoes more nuclear burning which produce more Na.

We see that a majority of the measured abundances of our objects lie above the maximum values, so while we do expect a significant Na enhancement based on the evolved CV models, they under-predict its magnitude compared to observations. There are several possible reasons for this discrepancy. Firstly, the rate of the $^{22}$Ne(p,$\gamma$)$^{23}$Na reaction have particularly large uncertainties compared to others in the NeNa cycle, stemming from contributions from possible low-energy resonances whose existence and properties have been debated in the literature (for further discussions on this topic, refer to \citealt{izzard_reaction_2007, iliadis_charged-particle_2010, kelly_new_2017}). JINA reaclib  V2.0 uses rates from \citet{iliadis_charged-particle_2010} (see their Table B.20). Fortunately, since the dependence of the yield of $^{23}$Na on the reaction rate is not linear and becomes relatively flat at fast rates as the $^{22}$Ne fuel gets used up, the magnitude of the uncertainty in the rate is not directly carried over to the yield (see \citealt{izzard_reaction_2007} for a more detailed explanation). Nevertheless, the $^{23}$Na abundance calculated by MESA and discussed throughout should only be taken to be accurate to within a factor of a few. 

Also, it is seen that irrespective of initial mass, the Na abundance plateaus at $P_{\rm orb} \lesssim 2$ h. This is likely to be the result of a nuclear statistical equilibrium being reached which prevents more Na from being produced in the outer layers than is initially produced at the core when it reaches the necessary temperature for the NeNa cycle to begin. This is further discussed in Appendix \ref{appendix:profiles} by looking at radial profiles of the abundances at several points in the evolution. 

Lastly, it is possible that our measured abundances are overestimated, which can occur if the donor temperatures are underestimated or if there is He enhancement that was unaccounted for in some objects, as discussed in Section \ref{ssec:he_enh} and Appendix \ref{appendix:He_enh}. Another possibility is that there are other physics that we have neglected that furthers the Na enhancement of evolved CVs. We explore one such process, namely rotational mixing, in the following section. 

\subsection{Rotational Mixing} \label{ssec:rotational_mixing}

Rotation effects in massive stars have been studied extensively in the literature (see \citealt{Maeder1998ASPC} for an overview of the various physical processes considered in models of rotating massive stars). 
Meanwhile, there has been comparatively less work done on this topic in lower mass stars. One reason for this is that the original treatment of mixing in rotating stars, the so-called Eddington-Sweet circulation, has a characteristic timescale $\propto \tau_{\rm KH}/\epsilon$ where $\tau_{\rm KH}$ is the Kelvin-Helmholtz timescale and $\epsilon$ is the ratio of centrifugal to gravitational acceleration \citep{Eddington1925Obs, mestel_mixing_1986}. This timescale is short and thus this process is very important for fast-rotating massive stars but less so for solar-type stars. However, more recent studies have shown that its effects may be significant even for lower mass stars \citep{Palacios2003A&A, chatzopoulos_effects_2012, istrate_models_2016}. In particular, mixing may change stellar abundances by allowing for advanced nuclear processing in a greater portion of the star, or by counteracting the effect of gravitational settling so that more heavier elements can be found in the outer layers (e.g. \citealt{istrate_models_2016} used MESA models to study this effect in proto-ELM WD binaries and found that it is important in their early stages). Furthermore, there can be contributions from other types of mixing such as the Spruit-Taylor magnetic diffusion (fluid motion resulting from toroidal magnetic field instabilities in differentially rotating stars \citep{tayler_adiabatic_1973, spruit_differential_1999}; e.g. \citealt{chatzopoulos_effects_2012} considered this mechanism in explaining carbon deficits observed in some compact object - solar-type binaries). 

Therefore, we also ran a few models including rotational mixing. This is implemented in MESA with am\_D\_mix\_factor (i.e. the rotational diffusion coefficient) which we set to the commonly used value of 1/30 from \citet{heger_presupernova_2000}. We also set D\_ES\_factor and D\_ST\_factor to one, corresponding to the inclusion of both the Eddington-Sweet and Spruit-Taylor mixing processes. 
As an in-depth exploration of rotational effects is beyond the scope of this paper, we just focus on a single comparison of models with mixing turned on and off. Because the timescale of main sequence evolution is highly sensitive to any changes in the internal structure of the stars, the inclusion of rotational mixing has a significant effect on how evolved the donor is when mass transfer begins. Thus, rather than comparing models with the same initial periods, with and without mixing, we should compare models that have similar trends in the effective temperature (i.e. are at similar evolutionary stages at fixed orbital period). 

\begin{figure*}
    \centering
    \includegraphics[width=0.8\textwidth]{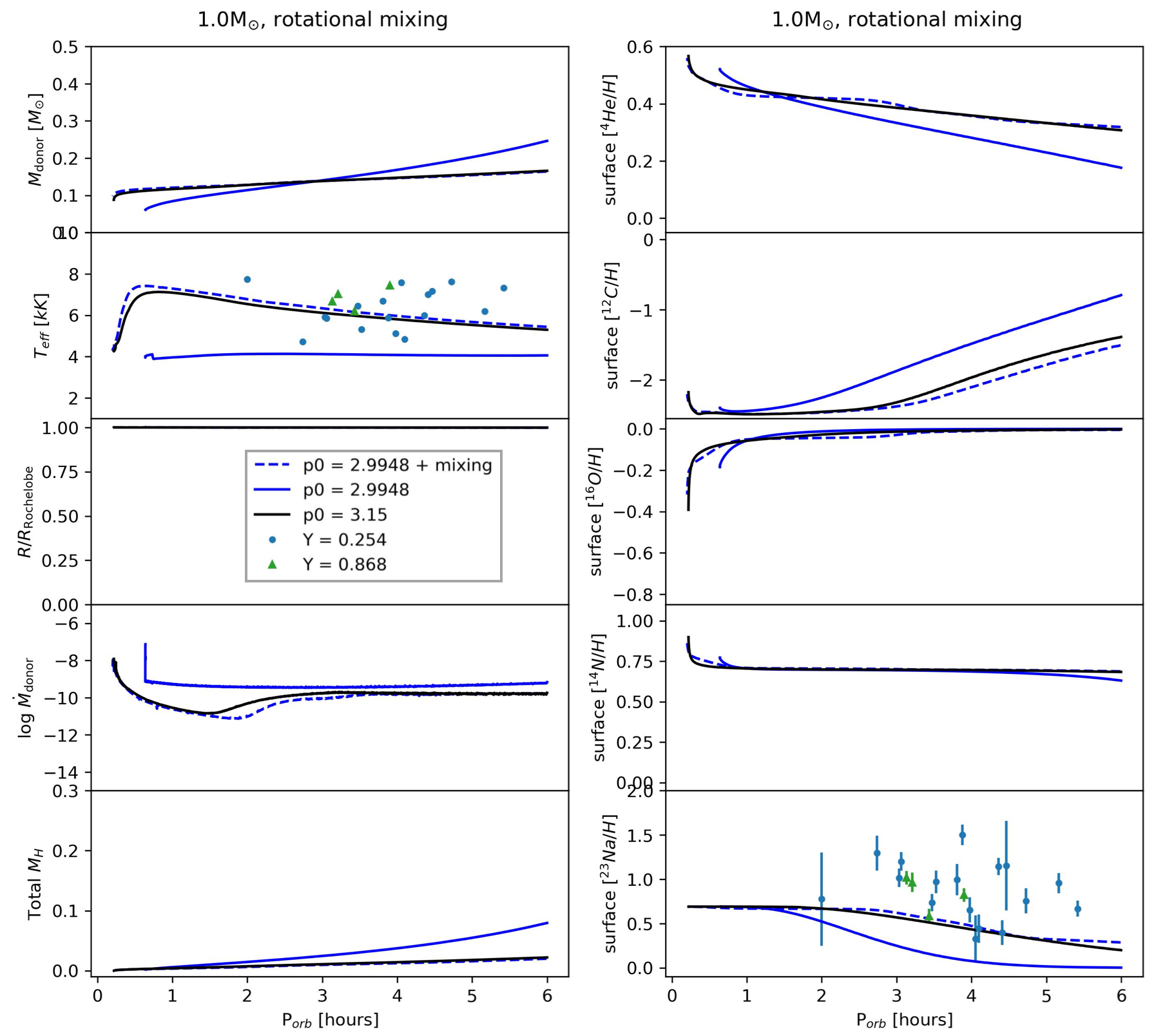}
    \caption{Results of MESA models for initial $M_{\rm donor} = 1.0 M_{\odot}$ with (blue dashed) and without (blue solid) rotational mixing for $p_0 = 2.9948$, as well as for $p_0 = 3.15$ without mixing (black solid). We see that the blue dashed and black solid lines have almost overlapping T$_{\rm eff}$ curves meaning that they are similarly evolved. Thus, at a given initial period, mixing causes the donor to become more evolved in a shorter period of time. We also find that there is very little difference in the black solid and blue dashed lines on the plot of [Na/H] on the right. Therefore, at a given evolutionary stage of the donor, mixing does not have a significant effect on the predicted Na abundance.}
    \label{fig:rotmixing}
\end{figure*}

Figure \ref{fig:rotmixing} plots the evolutionary tracks of reference (i.e. no mixing) models with initial periods $p_0 = 2.9948$ and $3.15$ (also plotted on Figures \ref{fig:physical} and \ref{fig:chemical}), as well as a model again with $p_0 = 2.9948$ but with mixing. From the panels on the left showing several basic donor parameters, we find that for a given initial period, the inclusion of mixing pushes the donor towards becoming more evolved in a shorter amount of time. In other words, the $T_{\rm eff}$ curves of the $p_0 = 2.9948$ model is raised upwards to almost overlap that of the $p_0 = 3.15$ model. 

Now comparing the plots of surface abundances to the right of the reference $p_0 = 3.15$ model against the $p_0 = 2.9948$ model with mixing, we see that there is little difference between them across all elements. In particular, there is no consistent rise in the Na abundance. Any small differences can be explained by the fact that they are not identically evolved and as they are smaller than most of the error bars in the measured abundances of our objects, such differences are not enough to account for the discrepancy between the models and observations. Thus, we find that given donors at a similar evolutionary stage, mixing does not have a significant effect on the surface abundances at the orbital periods of our interest. 

Moreover, we note that we did not include element diffusion processes (e.g. gravitational settling) which, as explored by \citet{istrate_models_2016}, work against rotational mixing, meaning that any impact on the abundances found here are likely optimistic. For completeness, we did run several models with element diffusion (implemented on MESA with \texttt{do\_element\_diffusion}) but found that once again, given donors that are similarly evolved, it has little effect on the predicted surface abundances. 

\subsection{Conclusions}

We have obtained follow-up high resolution spectroscopy of the 21 CVs with evolved donors first studied by \citetalias{el-badry_birth_2021}. The donors in these systems first overflowed their Roche lobes near the end of their main-sequence evolution and are thus predicted to have helium-enriched cores and anomalous surface abundances. 
Our goal was to measure the surface Na abundances of the donors to test for enhancement in Na from nuclear burning, which occured when the material now in the donors' photospheres was inside their convective cores. We also ran evolutionary models to find the range of initial parameters that lead to the formation of evolved CVs and test if they are able to explain the observed abundances. Our main findings are as follows:

\begin{enumerate}
  \item \textit{Na enhancement}: By measuring equivalent widths (EWs) of the Na 5900 \AA \ doublet from the observed spectra and comparing them to those of synthetic spectra generated over a grid of Na abundances (Figure \ref{fig:EW_regions}), we calculated [Na/H] values for the objects. As described in Section \ref{ssec:NLTE}, we also corrected for non-LTE effects. We find $0.3 \lesssim$ [Na/H] $\lesssim 1.5$ dex with a median value of 0.95 dex (i.e. nearly an order of magnitude enhancement compared to solar; Table \ref{tab:EWs_and_abun}).  Our study thus shows that surface Na enhancement, which has previously been detected in several CVs, is generically found in CVs with evolved donors.
  \item \textit{Evolutionary models for evolved CVs}: We ran binary models using MESA to investigate the initial binary parameters that produce evolved CVs similar to those in the observed sample. We modelled the WD as a point mass of 0.7 $M_{\odot}$ and tested initial donor masses of 1.0 and 1.5 $M_{\odot}$ over a range of initial periods (Section \ref{sec:mesa}). We found that for initial $M_{\rm donor} = 1.0 \ M_{\odot}$, initial periods of $3.0 \lesssim p_0 < 3.2$ days produced evolved donors with effective temperatures similar to the observed systems. For $M_{\rm donor} = 1.5 M_{\odot}$, the required period range was $0.7 \lesssim p_0 < 0.9$ days, which is shorter and narrower (Figure \ref{fig:physical}).
  \item \textit{Na enhancement predicted by MESA models}: Models of evolved CVs predict a significant enhancement in the surface Na abundance, consistent with our observations. At the orbital periods of the observed systems, the predicted amount of enhancement is sensitive to both the initial period and initial donor mass. The ubiquitous Na enhancement of evolved CV donors can thus likely be understood as a consequence of nuclear processing in the donors during their main-sequence evolution. Models with higher initial donor masses generally predict stronger Na enhancement, while models for ``normal'' unevolved CVs predict no enhancement (Figure \ref{fig:chemical}). 
  \item  \textit{Connection to Future Modeling}: A majority of the observed donors have higher inferred Na abundances than predicted by the MESA models. Possible sources of this discrepancy include the large uncertainty in the reaction rate of the step producing $^{23}$Na in the NeNa cycle, or missing physical processes (mixing and diffusion are described below). These may be worth further investigation when constructing future models. The discrepancy could also be due to an overestimation of the calculated abundances which could result from an underestimation of the donor effective temperature or unaccounted for He enhancement (see Section \ref{ssec:he_enh} and Appendix \ref{appendix:He_enh}). 
  \item  \textit{Effects of rotational mixing}: We explored the possibility of additional enhancement through rotational mixing by turning on Eddington-Sweet circulation and Spruit-Taylor instabilities in our MESA models. We found that at a given initial donor mass, mixing pushed the donor to become evolved more quickly. However, given donors at the same evolutionary stage (i.e. similar curves of effective temperature at a given period), mixing had very little effect on the predicted Na abundances (Section \ref{ssec:rotational_mixing}). Therefore, while mixing does have an effect on the evolutionary path of a system given the same initial conditions, it is unlikely to be able to account for the extra Na enhancement seen in our objects.
\end{enumerate}

\section*{Acknowledgements}

We thank Shrinivas Kulkarni for help acquiring observational data, and John Thorstensen, Tom Marsh, and the ZTF stellar variables group for helpful discussions. We also thank the anonymous referee for the detailed feedback. 

The data presented herein were obtained at the W. M. Keck Observatory, which is operated as a scientific partnership among the California Institute of Technology, the University of California and the National Aeronautics and Space Administration. The Observatory was made possible by the generous financial support of the W. M. Keck Foundation.

The authors wish to recognize and acknowledge the very significant cultural role and reverence that the summit of Maunakea has always had within the indigenous Hawaiian community.  We are most fortunate to have the opportunity to conduct observations from this mountain.

Based on observations obtained with the Samuel Oschin Telescope 48-inch and the 60-inch Telescope at the Palomar
Observatory as part of the Zwicky Transient Facility project. ZTF is supported by the National Science Foundation under Grant
No. AST-2034437 and a collaboration including Caltech, IPAC, the Weizmann Institute for Science, the Oskar Klein Center at
Stockholm University, the University of Maryland, Deutsches Elektronen-Synchrotron and Humboldt University, the TANGO
Consortium of Taiwan, the University of Wisconsin at Milwaukee, Trinity College Dublin, Lawrence Livermore National
Laboratories, and IN2P3, France. Operations are conducted by COO, IPAC, and UW.

\section*{Data availability}
The data underlying this article are available upon reasonable request to the corresponding author. 

\bibliographystyle{mnras}



\appendix

\section{Observing log} \label{appendix:log}

Table \ref{appendix:log} summarizes the observing dates and instrument specifications for the spectra collected for all objects. Note that the ``ID''s used here are internal to \citetalias{el-badry_birth_2021} and this work -- the ``P'' stands for period, followed by a number which is the period in hours, and the letter can be used to differentiate objects with the same period to two decimal places.

We observed 20 (out of 21) systems with the Echellette Spectrograph and Imager (ESI) in Echellette mode on the Keck II telescope. The majority (18) of these were observed on 09/02/22 and 10/24/22. For these, the wavelength coverage was between 4000 - 10150 \AA, divided across 10 orders. During the first night, 300s exposures were taken using the 0.3" slit, which resulted in a median resolution of R $\sim$ 11700. The resolutions were calculated using the [OI]5577\AA \ sky line by taking the ratio of the central wavelength to the FWHM of the line obtained from Gaussian fitting. Signal-to-noise ratios (SNRs) were calculated over 20 pixels centered around 5850 and 5890 \AA \, surrounding the Na doublet, by taking the ratio of the mean flux over the mean flux error. This gave us a median SNR (averaged over the two regions) of $\sim 6$ (the values for each spectra can be found in Table \ref{tab:log}). Based off of these SNRs, and checking the effect of orbital smearing in determining the rotational velocity was minimal over longer exposure times, it was decided that for the next night, 600s exposures with a 0.5" slit would be used instead, which gave us R $\sim$ 9300 and SNR $\sim$ 21. The remaining two objects with ESI spectra were observed by a collaborator as backup targets on 03/06/22 and multiple 300s exposures were taken with the 1.0" slit, resulting in R $\sim$ 4800. To improve the SNR, these spectra were co-added after correcting for each of their radial velocities which gave us SNRs $\sim$ 17.

As described in Section \ref{sec:esi}, one object (P\_2.74a) was observed using LRIS with 90s exposures over its complete orbit.

  \begin{table*} 
     \centering
     \begin{tabular}{|c c c c c c c c|}
          \hline
         ID & Date [UTC] & MJD & Exposure [s] & Slit width ["] & Resolution & SNR & Instrument   \\
          \hline
         P\_2.00a & 09/02/2022 & 59732.26710000 & 300 & 0.3 & 11720 & 5.77 & ESI \\
         P\_2.74a & 06/01/2022 & 59877.60220858 & 63x(r), 78x(b) 90 & 1.0 & 1543 & 34.92 & LRIS \\
         P\_3.03a & 10/24/2022 & 59877.21684573 & 600 & 0.5 & 9352 & 11.26 & ESI \\
         P\_3.06a & 10/24/2022 & 59825.40700344 & 600 & 0.5 & 9332 & 13.25 & ESI \\
         P\_3.13a & 10/24/2022 & 59877.31297247 & 600 & 0.5 & 9321 & 28.81 & ESI \\
         P\_3.21a & 10/24/2022 & 59825.53532983 & 600 & 0.5 & 9299 & 9.20 & ESI \\
         P\_3.43a & 10/24/2022 & 59877.43533115 & 600 & 0.5 & 9473 & 18.65 & ESI \\
         P\_3.48a & 03/06/2022 & 59825.54595483 & 3x300 & 1.0 & 4845 & 21.03 & ESI \\
         P\_3.53a & 10/24/2022 & 59877.54255511 & 600 & 0.5 & 9360 & 22.71 & ESI \\
         P\_3.81a & 09/02/2022 & 59644.55657553 & 300 & 0.3 & 11611 & 6.50 & ESI \\
         P\_3.88a & 03/06/2022 & 59877.32205094 & 3x300 & 1.0 & 4825 & 13.73 & ESI \\
         P\_3.90a & 10/24/2022 & 59825.44070830 & 600 & 0.5 & 9321 & 38.95 & ESI \\
         P\_3.98a & 10/24/2022 & 59644.57136847 & 600 & 0.5 & 9342 & 14.60 & ESI \\
         P\_4.06a & 09/02/2022 & 59877.61201310 & 300 & 0.3 & 11459 & 16.39 & ESI \\
         P\_4.10a & 10/24/2022 & 59877.55295580 & 600 & 0.5 & 9321 & 28.79 & ESI \\
         P\_4.36a & 10/24/2022 & 59877.63229504 & 600 & 0.5 & 9277 & 13.32 & ESI \\
         P\_4.41a & 10/24/2022 & 59877.30323879 & 600 & 0.5 & 9322 & 21.04 & ESI \\
         P\_4.47a & 09/02/2022 & 59825.45260830 & 300 & 0.3 & 12214 & 2.38 & ESI \\
         P\_4.73a & 10/24/2022 & 59877.33081067 & 600 & 0.5 & 9299 & 23.49 & ESI \\
         P\_5.17a & 09/02/2022 & 59877.33934018 & 300 & 0.3 & 11725 & 9.73 & ESI \\
         P\_5.42a & 10/24/2022 & 59877.53413636 & 600 & 0.5 & 9318 & 38.58 & ESI \\
          \hline  
     \end{tabular}
     \caption{An observing log for all objects. For clarity, the date provided is the date in UTC at the start of the observing run. The resolutions were calculated using the FWHM from the Gaussian fitting of the [OI]5577\AA \ line in the sky spectra. The signal-to-noise ratio (SNR) values provided are the averages between the values calculated over 20 pixels around 5890 and 5850 \AA, above and below the Na doublet (for ESI spectra, the values listed here are of the sixth order - those of the fifth order are comparable). For all ESI observations, Echellete mode was used and for LRIS, both the red and blue spectra were taken.}
     \label{tab:log}
 \end{table*}

\section{H-alpha lines} \label{appendix:h-alpha}

\begin{figure*}
    \centering
    \includegraphics[width=0.8\textwidth]{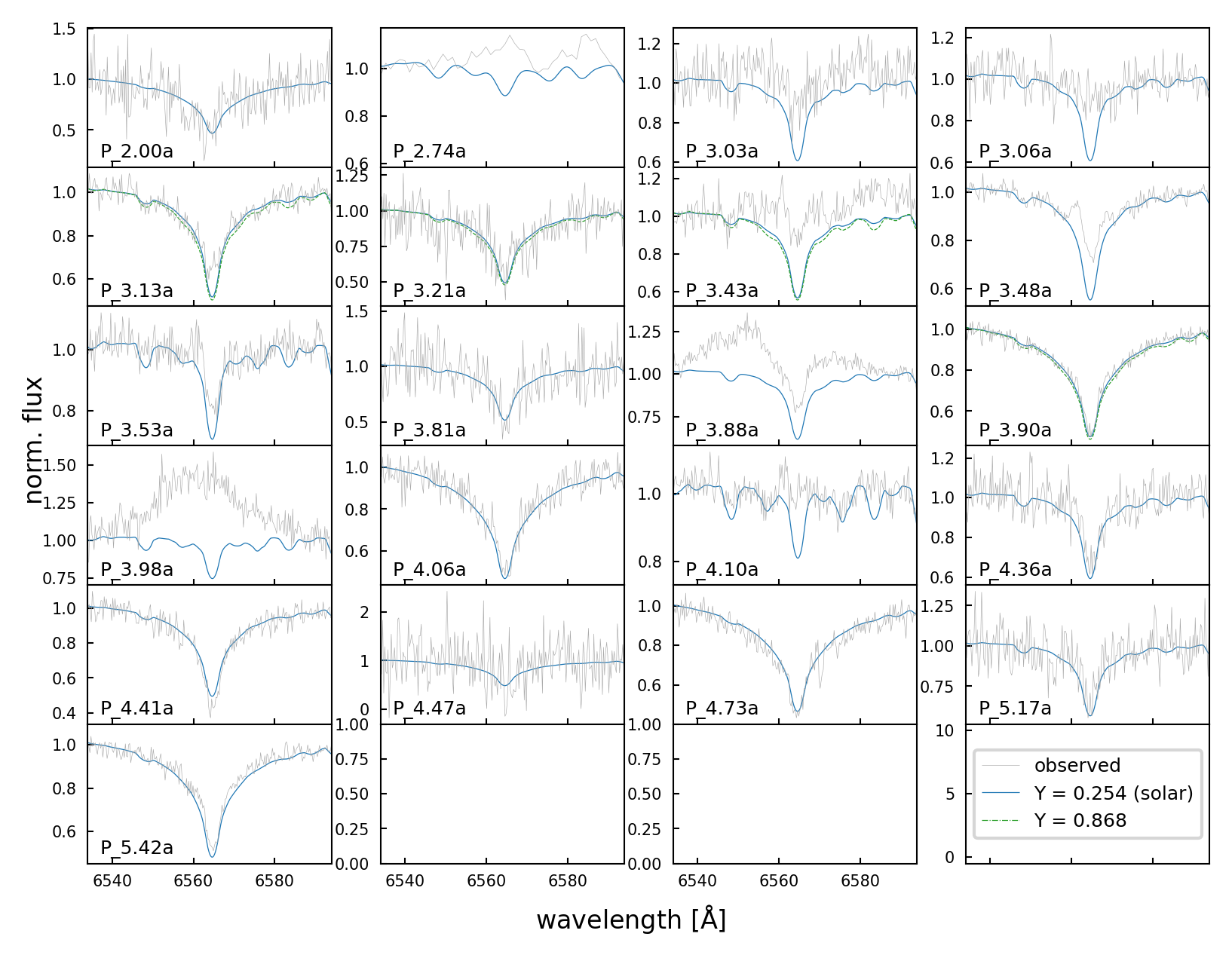}
    \caption{The spectra at the H$\alpha$ transition for all objects, plotted with their corresponding model spectra. It is seen that for about half of the objects, the line of the observed spectra is visibly shallower than that of the model or is an emission instead of an absorption. These objects are likely mass-transferring with contribution from an accretion disk.}
    \label{fig:h_alpha}
\end{figure*}

We plot H-$\alpha$ lines for all objects in Figure \ref{fig:h_alpha}, along with the Kurucz model spectra with solar abundances at the corresponding temperatures. It is seen that the model provides a relatively good fit to the data for around half of the objects. Meanwhile, several of them have noticeably shallower observed absorption lines compared to their models (e.g. P\_3.03a, P\_3.06a, P\_3.43a, P\_3.53a) while others have little to no absorption or instead show emission (e.g. P\_2.74a, P\_3.98, P\_4.10a). The same features were identified for many of the same objects in \citetalias{el-badry_birth_2021} and can most likely be explained by the contribution from an accretion in systems with ongoing accretion. 

\section{Interstellar lines} \label{appendix:IS}

\begin{figure*}
    \centering
    \includegraphics[width=0.7\textwidth]{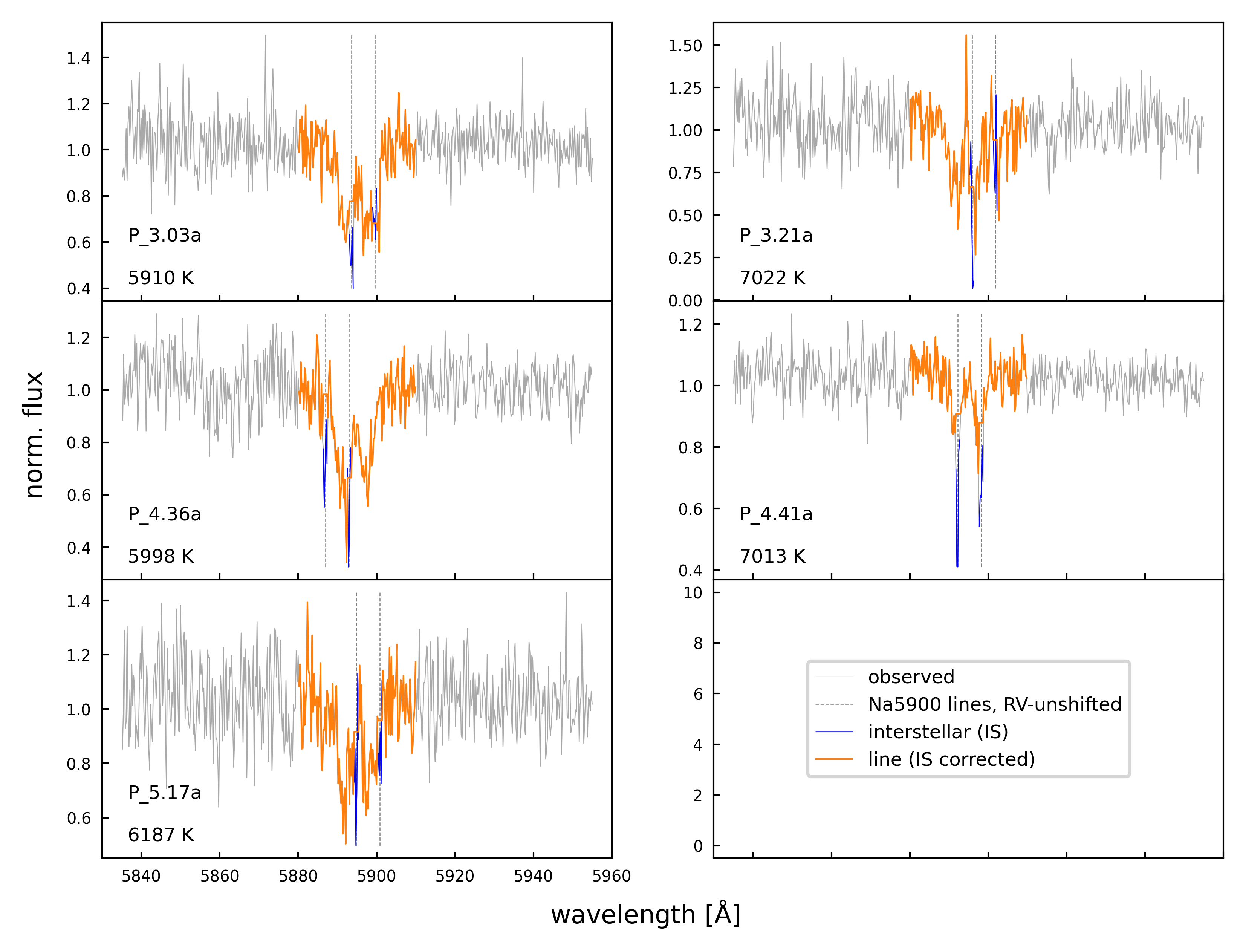}
    \caption{The Na 5900 \AA \ doublet for several representative objects showing the removal of interstellar (IS) lines. The IS lines can be distinguished from the donor lines as they are not broadened and observed very closely to their rest wavelengths (i.e. not shifted with the RV of the donor).}
    \label{fig:IS_lines}
\end{figure*}

The plots in Figure \ref{fig:IS_lines} show interstellar Na 5900\AA \ doublet lines highlighted in blue which are removed in the calculation of equivalent widths, using the spectra in orange. This can be done because interstellar lines are not shifted by the radial velocity of the star and so are observed very closely to their rest wavelengths. While for many objects, they are relatively narrow and indistinguishable from noise, in cases like P\_4.41a, their contribution can be significant and thus important to account for.  

\section{Contribution from a circumbinary disk} \label{appendix:circumbinary_disk}

While we have masked out narrow interstellar lines, one may reasonably ask whether there could potentially be contribution from a circumbinary disk. Lines from such a disk would be broadened due to its own orbital velocity around the binary system and thus would be more challenging to distinguish from the lines of the donor. 

\begin{figure}
    \centering
    \includegraphics[width=0.35\textwidth]{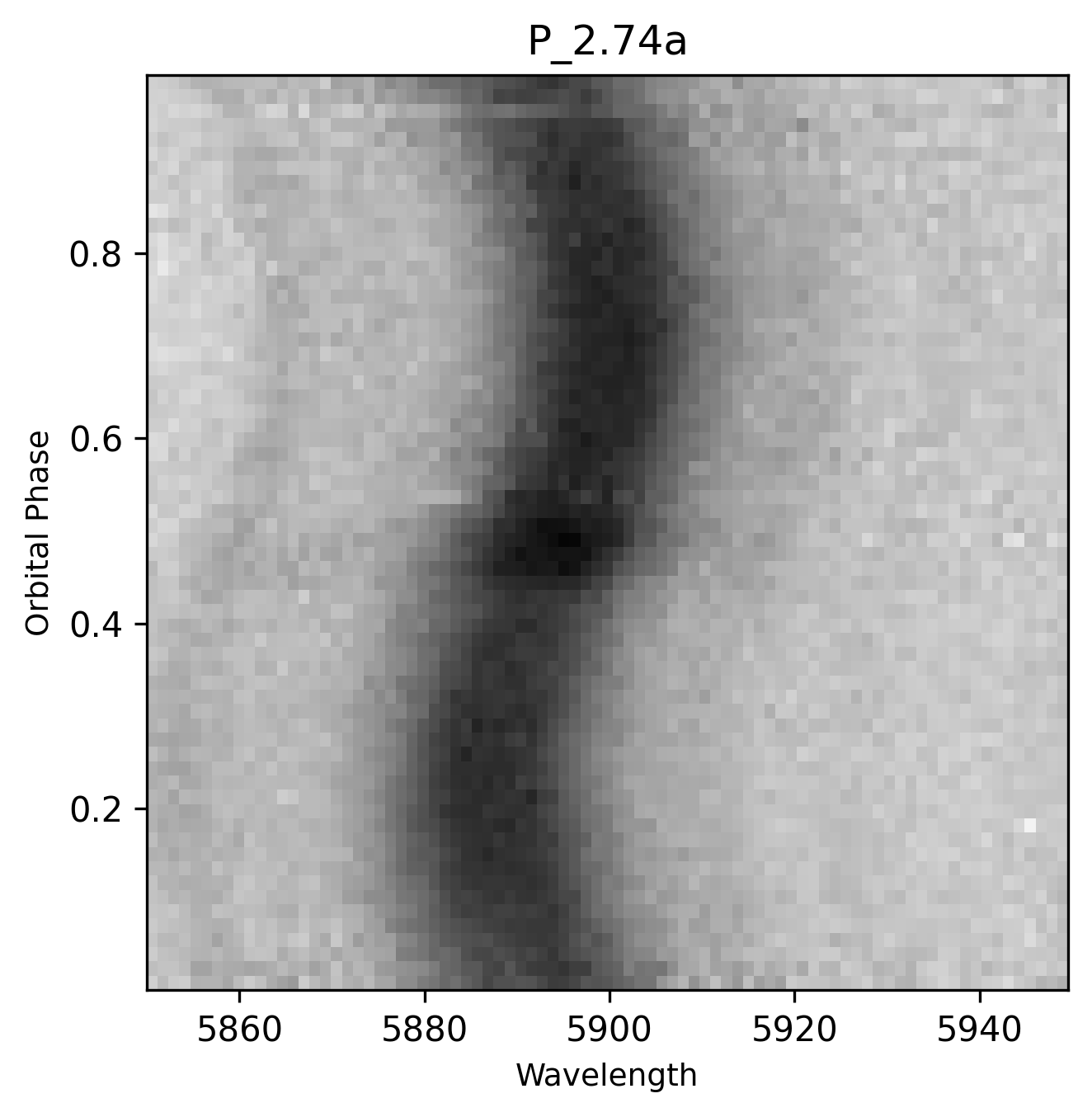}
    \caption{A trailed spectrum showing the Na doublet (unresolved at this resolution) for P\_2.74a collected over one period. An eclipse of the WD accretor occurs at phase = 0.5. The line is close to sinusoidal and follows the trend of neighbouring features, meaning it is tracing the donor without significant contribution from an external source such as a circumbinary disk.}
    \label{fig:river}
\end{figure}

One way to check whether or not this is the case is to obtain multiple spectra over the course of a complete orbit and trace radial velocity shifts of the line. This was done for one object in our sample, P\_2.74a, for which we obtained 90s exposures over 2.75 hours using LRIS (Section \ref{sec:esi} and Table \ref{appendix:log}) . Calculating the orbital period using the MJD at mid-exposure for each spectrum, stacking all of the normalized spectra, and displaying them as an image, we get a trailed spectrum as shown in Figure \ref{fig:river}, zoomed into the Na doublet (note that the two lines cannot be resolved with LRIS so they are seen as a single dark band on the plot). Qualitatively, it is seen that the line moves back and forth along with any neighbouring features as it traces the donor's orbital motion. Thus, there is no clear sign of contamination from ex. a circumbinary disk which will add a separate component to the radial velocity of the donor and cause deviation from a sinusoidal evolution over the orbit. While this was only done for a single object, it does show that significant Na enhancement can be present without evidence of a circumbinary disk. 

\section{Abundances with He enhancement} \label{appendix:He_enh}

If He enhancement is unaccounted for, it generally results in an overestimation of the Na abundance. Thus we calculate the inferred [Na/H] of all objects by comparison to models with a He enhanced mass fraction of $Y = 0.569$. The results are summarized in Table \ref{tab:abun_he_0p5}, where we see that with the exception of P\_2.74a, all other objects see a decrease in the calculated [Na/H] compared to those from Table \ref{tab:EWs_and_abun} (at the low temperature of P\_2.74a, the He enhanced model has a steeper curve of growth so that it predicts lower Na abundances compared to the solar model for small EWs but eventually crosses over). In this case, [Na/H] ranges $\sim$ 0.2 to 1.5 dex with a median value of 0.826 dex. Therefore, while there is a decrease in the inferred Na enhancement if He enhancement is present for all objects, it remains quite significant. 

\begin{table}
    \centering
    \begin{tabular}{|c c c|}
         \hline
         ID & [Na/H] (Y = 0.254) & [Na/H] (Y = 0.569) \\
         \hline
         P\_2.00a & 0.693 $\pm$ 0.517 & 0.597 $\pm$ 0.506 \\
         P\_2.74a & 1.296 $\pm$ 0.197 & 1.455 $\pm$ 0.179 \\
         P\_3.03a & 1.016 $\pm$ 0.105 & 0.852 $\pm$ 0.105 \\
         P\_3.06a & 1.201 $\pm$ 0.104 & 1.033 $\pm$ 0.099 \\
         P\_3.13a * & 1.016 $\pm$ 0.076 & 1.016 $\pm$ 0.076 \\
         P\_3.21a * & 0.964 $\pm$ 0.108  & 0.964 $\pm$ 0.108 \\
         P\_3.43a * & 0.583 $\pm$ 0.085 & 0.583 $\pm$ 0.085 \\
         P\_3.48a & 0.736 $\pm$ 0.096 & 0.227 $\pm$ 0.086 \\
         P\_3.53a & 0.970 $\pm$ 0.127 & 0.792 $\pm$ 0.127 \\
         P\_3.81a & 0.995 $\pm$ 0.177 & 0.826 $\pm$ 0.173 \\
         P\_3.88a & 1.502 $\pm$ 0.116 & 1.305 $\pm$ 0.106 \\
         P\_3.90a * & 0.822 $\pm$ 0.078 & 0.822 $\pm$ 0.078 \\
         P\_3.98a & 0.653 $\pm$ 0.142 & 0.475 $\pm$ 0.141 \\
         P\_4.06a & 0.328 $\pm$ 0.265 & 0.181 $\pm$ 0.251 \\
         P\_4.10a & 0.442 $\pm$ 0.162 & 0.260 $\pm$ 0.162 \\
         P\_4.36a & 1.143 $\pm$ 0.098 & 0.978 $\pm$ 0.095 \\
         P\_4.41a & 0.397 $\pm$ 0.141 & 0.237 $\pm$ 0.137 \\
         P\_4.47a & 1.152 $\pm$ 0.506 & 0.973 $\pm$ 0.525 \\
         P\_4.73a & 0.744 $\pm$ 0.143 & 0.595 $\pm$ 0.139 \\
         P\_5.17a & 0.956 $\pm$ 0.115 & 0.795 $\pm$ 0.113 \\
         P\_5.42a & 0.668 $\pm$ 0.091 & 0.505 $\pm$ 0.089 \\
         \hline
    \end{tabular}
    \caption{Na abundances calculated using He enhanced models with $Y = 0.569$. For ease of comparison, we also add a column from Table \ref{tab:EWs_and_abun} using solar abundance ($Y = 0.256$) models of He. The objects marked with an asterisk (*) are those whose abundances are calculated using $Y = 0.868$ models as discussed in Section \ref{ssec:he_enh}. All abundances here have been corrected for NLTE effects (Section \ref{ssec:NLTE}). With the exception of P\_2.74a, [Na/H] for all other objects decrease with increasing He abundance. The values now range from $\sim$ 0.2 - 1.5 dex with the median being 0.826 dex which is however, still a significant Na enhancement.}
    \label{tab:abun_he_0p5}
\end{table}

\section{MESA profile plots} \label{appendix:profiles}

\begin{figure*}
    \centering
    \includegraphics[width=0.65\textwidth]{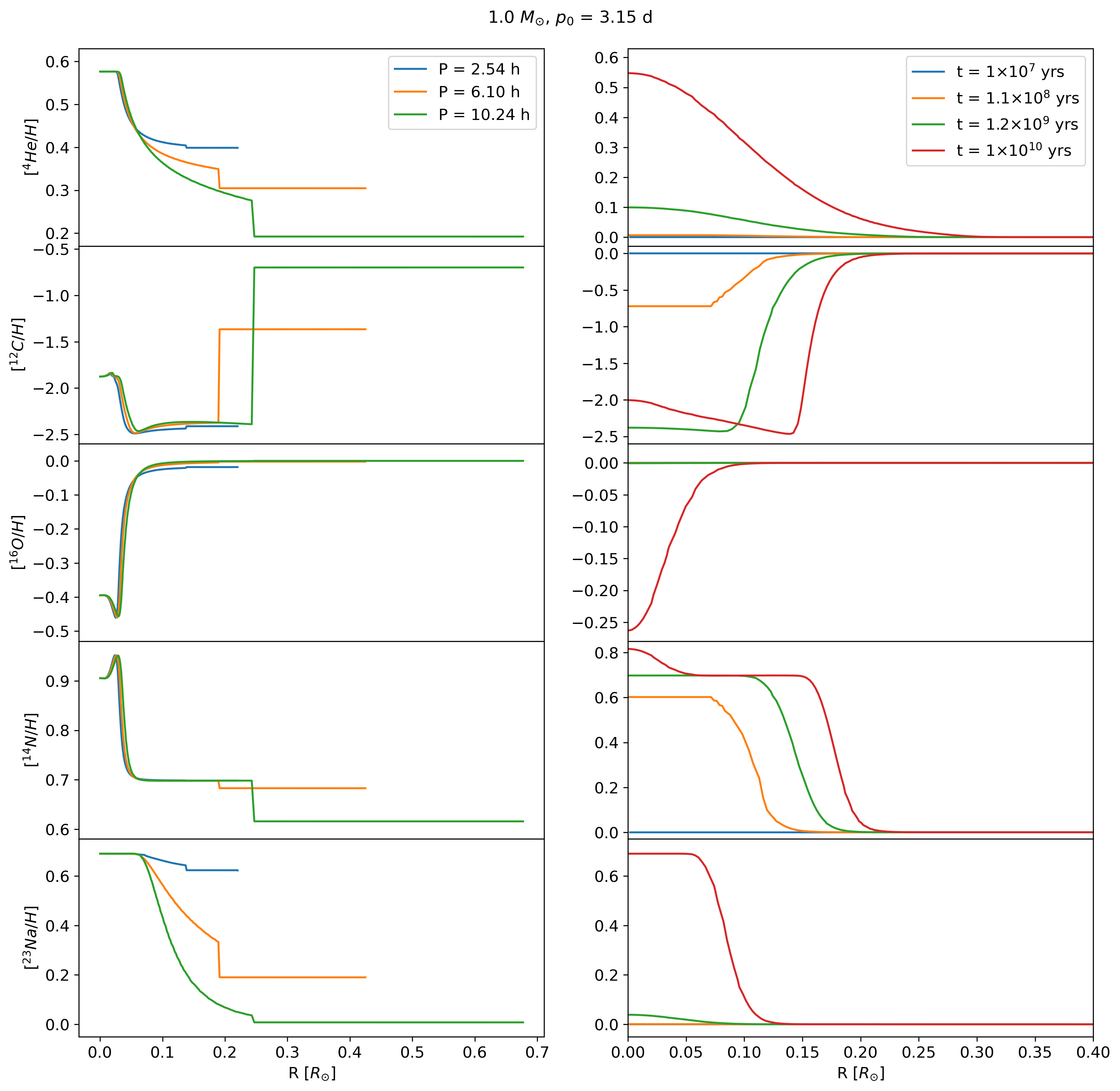}
    \caption{Profiles of the MESA models for initial parameters $M_{\rm donor} = 1.0 M_{\odot}$ and $p_0 = 3.15$ days, showing the abundances of several elements at several different orbital periods (left) and star ages (right). The plots on the left show that as the donor evolves towards shorter periods, its radius decreases as it sheds more of its envelope and the surface Na abundance increases. Meanwhile, from the plots on the right (spanning a much greater length of time), we find that the central Na abundance increases significantly (from $\sim$ 0.05 to 0.7 dex) between $10^9$ to $10^{10}$ years where it is maintained thereafter.}
    \label{fig:mesa_profiles}
\end{figure*}

Figure \ref{fig:mesa_profiles} show several profiles (plots as a function of the radial coordinate of the donor at some snapshots during the evolution) of the chemical abundances for the model with initial donor mass of $1.0 M_{\odot}$ and initial orbital period of 3.15 days. 

The plots on the left show profiles at three orbital periods $\lesssim 10$ hours. As expected, the radius of the donor falls as the period decreases and more of its mass is transferred to the WD. More importantly, the chemical abundances as a function of the fraction of radius evolves over time. In particular, it is seen that at the longest period (green), there is a core rich in Na but its abundance falls off rapidly to solar values at the surface. Therefore, if the donor is observed at this period, no Na enhancement will be observed. As the donor continues to evolve, it loses its outer envelope and undergoes further burning so that we begin to see a rise in Na abundance at the outer layers. If the donor is observed at the shortest period plotted (blue), we will observe a significant Na enhancement of $\sim$ 0.5 dex. Note that the abundances at the central region ($\lesssim 0.1 R_{\odot}$) plateaus to a constant value at all periods and in particular for Na, the abundance at the outer layers closely approach but never exceeds this value, placing a limit onto the achievable enhancement. This points to an equilibrium being reached between the two regions. 

To identify when the core abundance was established, we also provide profiles spanning a much greater length of time, from $\sim$ 10 Myr - 10 Gyr, on the right. We see that there is a large increase in the central Na abundance between 1 Gyr to 10 Gyr as it has had a long time to build up, where it increases from $\sim$ 0.05 to the maximum of 0.7 dex. The small rise in central temperature between these two times may also contribute slightly as it can result in a significant increase in the reaction rate of the step producing the Na. 


\bsp	
\label{lastpage}
\end{document}